\pdfoutput=1
\PassOptionsToPackage{numbers,sort&compress}{natbib}
\documentclass[preprint,12pt]{elsarticle}  

\usepackage[T1]{fontenc}
\usepackage[utf8]{inputenc}
\usepackage{lmodern}
\usepackage{microtype}

\usepackage{amsmath,amssymb}
\usepackage{algorithmic}
\usepackage[ruled,vlined]{algorithm2e}
\usepackage[a4paper, margin=1in]{geometry}
\usepackage{tabularx}
\usepackage{booktabs}
\usepackage{ragged2e}
\usepackage{graphicx}
\makeatletter
\def\ps@pprintTitle{%
  \let\@oddhead\@empty
  \let\@evenhead\@empty
  \let\@oddfoot\@empty
  \let\@evenfoot\@oddfoot}
\makeatother

\usepackage[numbers,sort&compress]{natbib}
\usepackage{hyperref}
\hypersetup{
  colorlinks=true,
  linkcolor=blue,
  citecolor=blue,
  urlcolor=blue,
  pdftitle={Explainable AI-Enhanced Supervisory Control for High-Precision Spacecraft Formation},
  pdfauthor={Reza Pirayeshshirazinezhad}
}

\begin{document}
\begin{frontmatter}

\title{Explainable AI-Enhanced Supervisory Control for High-Precision Spacecraft Formation}
\author{Reza Pirayeshshirazinezhad}
\address{Texas A\&M University, TX, USA}

\begin{abstract}

We use artificial intelligence (AI) and supervisory adaptive
control systems to plan and optimize the mission of precise spacecraft formation. Machine learning and robust
control enhance the efficiency of spacecraft precision formation of the Virtual Telescope for X-ray Observation
(VTXO) space mission. VTXO is a precise formation of two separate spacecraft making a virtual telescope
with a one-kilometer focal length. One spacecraft carries the lens and the other spacecraft holds the camera
to observe high-energy space objects in the X-ray domain with 55 milli-arcsecond angular resolution accuracy. Timed automata for supervisory control, Monte
Carlo simulations for stability and robustness evaluation, and integration of deep neural networks for optimal estimation of
mission parameters, satisfy the high precision mission criteria. We integrate deep neural networks with a constrained, non-convex dynamic optimization pipeline to predict optimal mission parameters, ensuring precision mission criteria are met. AI framework provides explainability by predicting the resulting energy consumption and mission error for a given set of mission parameters. It allows for transparent, justifiable, and real-time trade-offs, a capability not present in traditional adaptive controllers. The results show reductions in energy consumption and improved
mission accuracy, demonstrating the capability of the system to address dynamic uncertainties and disturbances.
\end{abstract}

\begin{graphicalabstract}
\begin{figure}[ht!]
\centering
\raggedright
\includegraphics[width=0.05\linewidth]{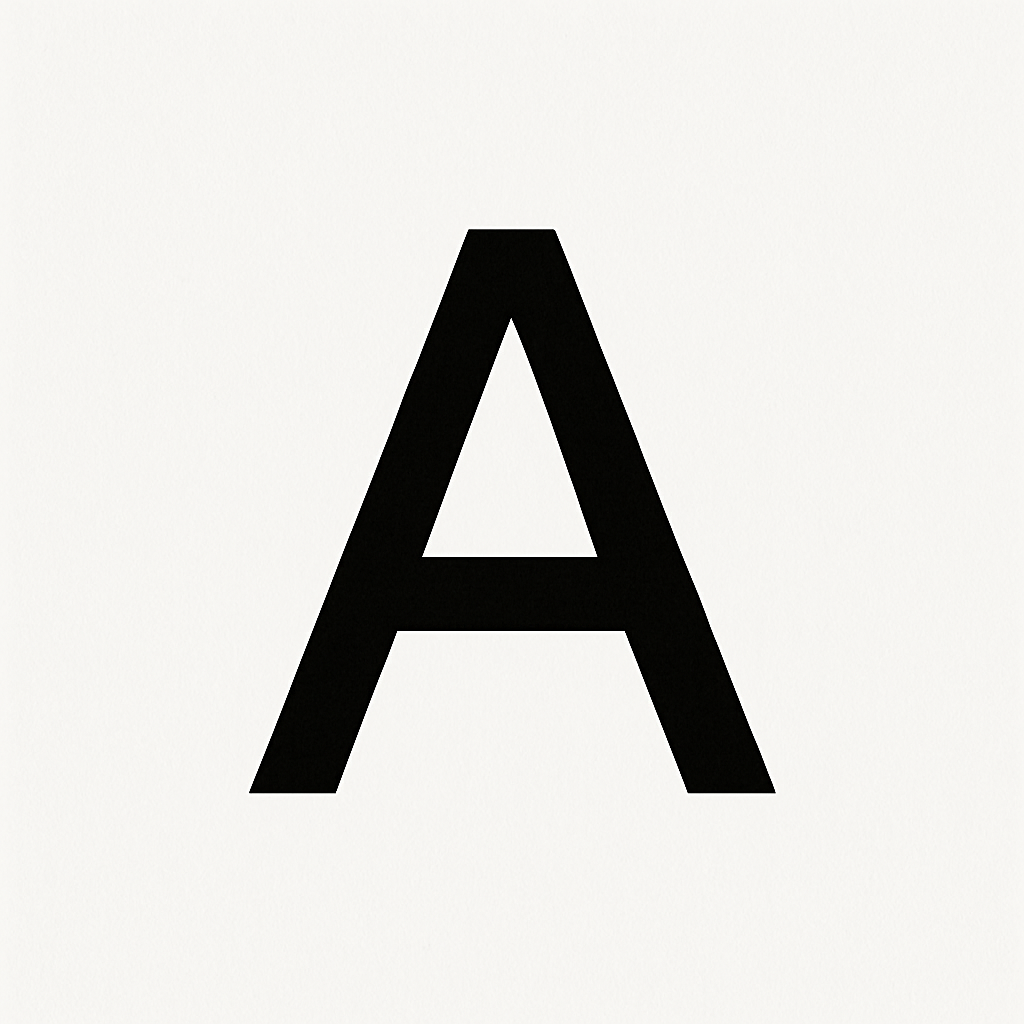}
\includegraphics[width=1\linewidth]{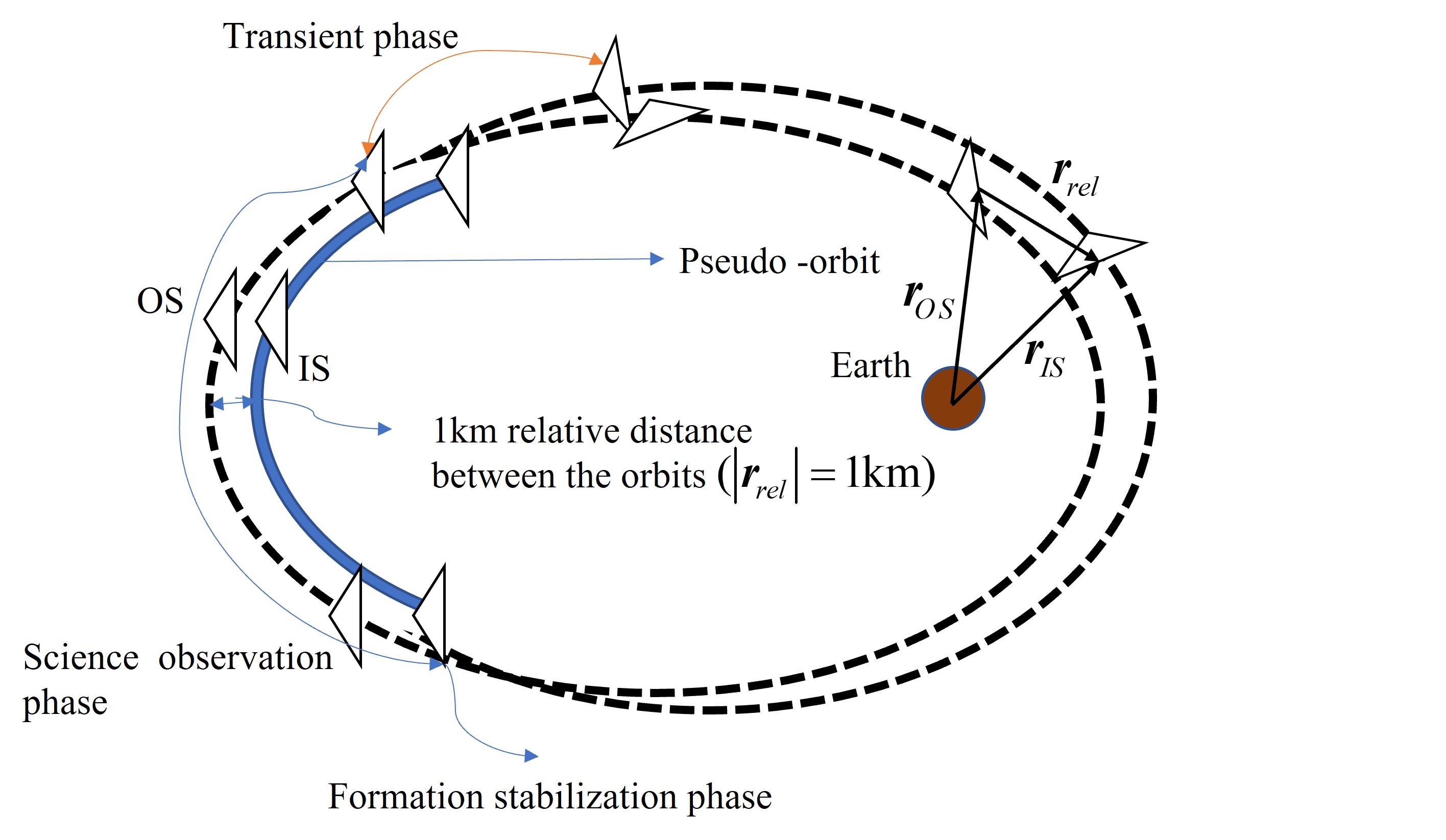}
\raggedright
\includegraphics[width=0.05\linewidth]{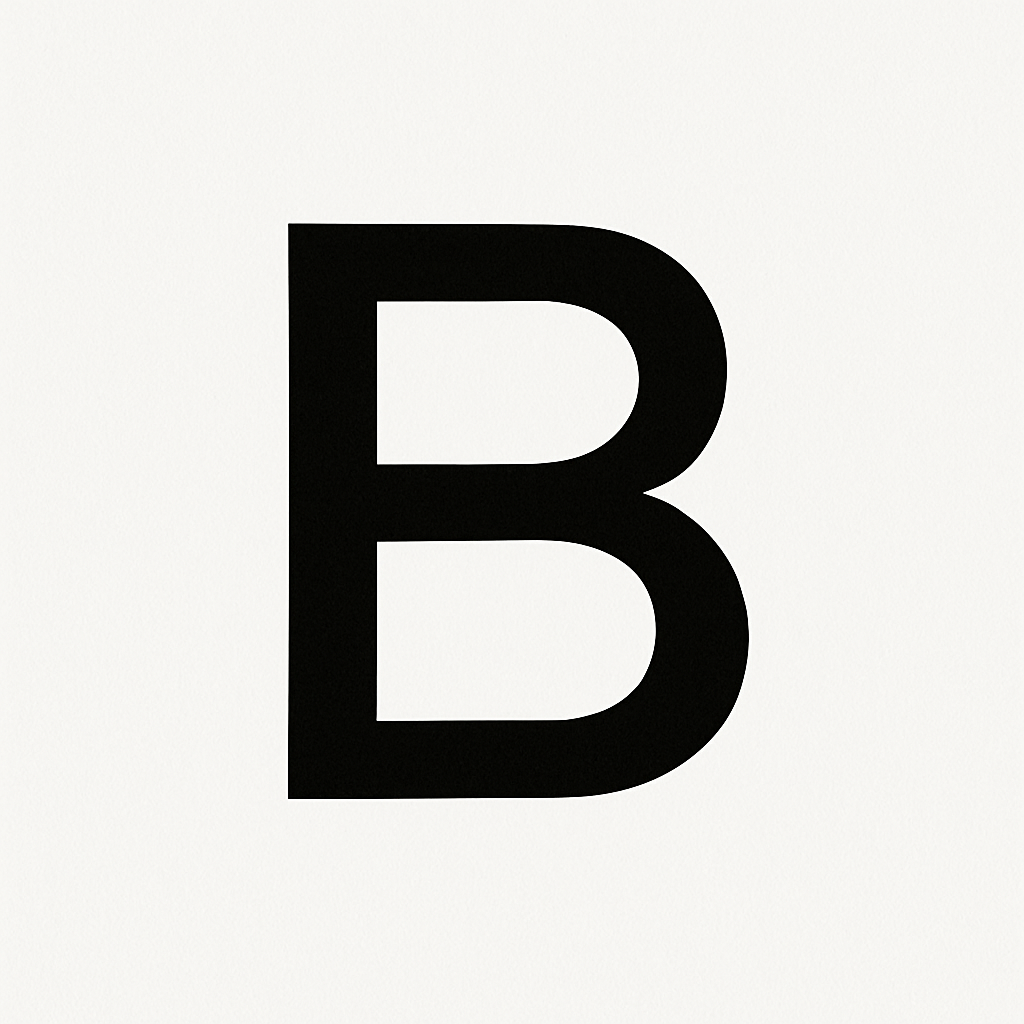}
\vspace{0.3cm}
\includegraphics[width=0.9\linewidth]{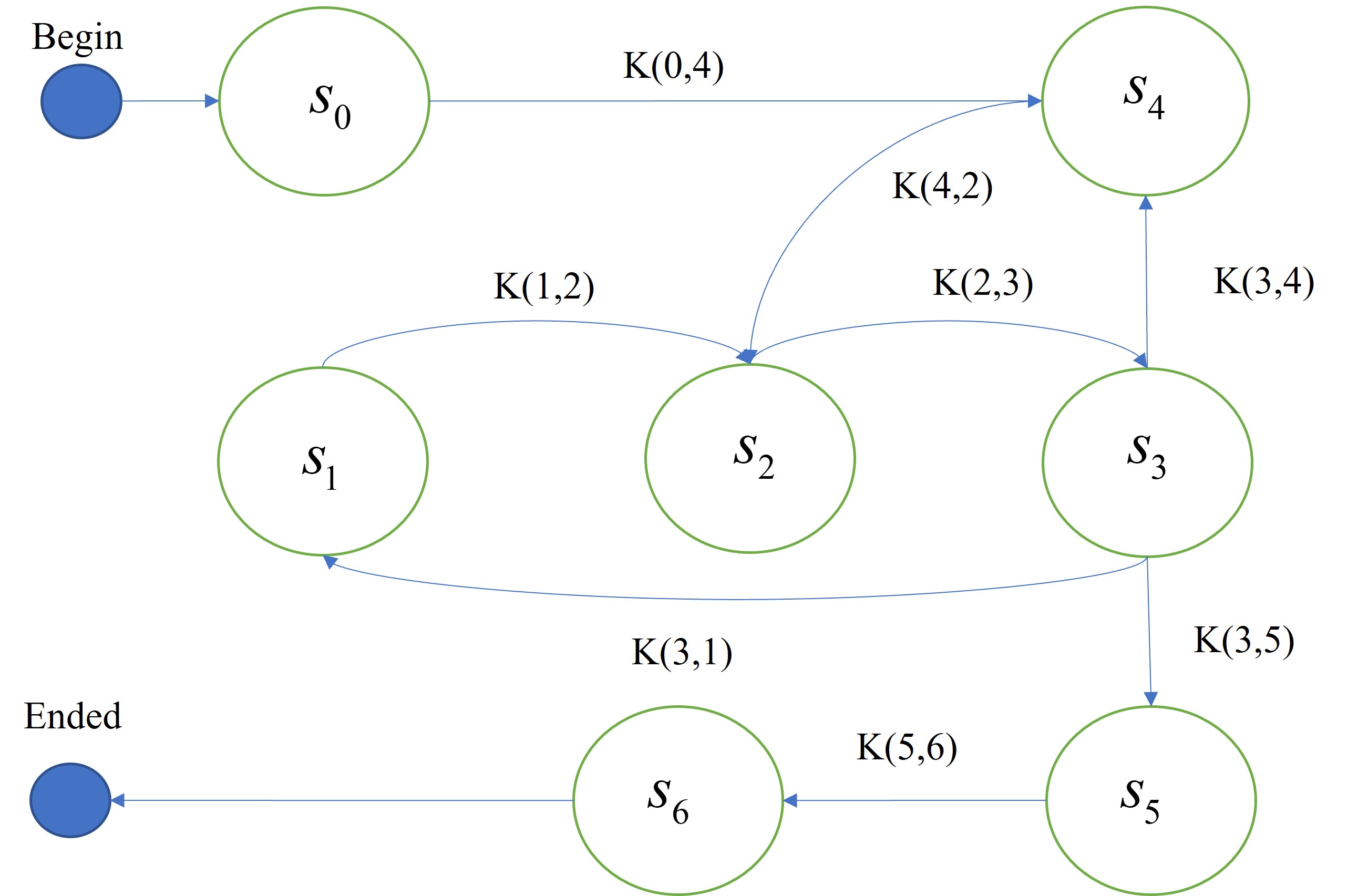}
\caption{(A) The orbits and ACS phases in the formation are shown. The pseudo-orbit keeps the formation for $|\pmb{r}_{rel}|=1\text{ km}$ with 1 meter accuracy. $\pmb{r}_{rel}=\pmb{r}_{F}-\pmb{r}_{L}$. $\pmb{r}_{L}$ and $\pmb{r}_{F}$ show the distance vectors of the spacecraft.Supervisory adaptive control system. (B) $\pmb{s}$ represents the set of discrete states of finite automata, and $\pmb{k}$ represents the set of all transitions in finite automata. Begin and Ended represent the beginning and ending of VTXO mission.}
\label{Orbits}
\end{figure}

\end{graphicalabstract}

\begin{highlights}
\item Introduces a timed automata-based supervisory control framework for spacecraft formation flying.
\item Introduces sliding mode control for high-precision relative position formation.
\item Introduces an explainable AI for performance prediction over the optimal mission parameters.
\item Introduces a neural network based solver for a unified constrained, non-convex dynamic optimization.
\end{highlights}


\end{frontmatter}

\section{Introduction}
\label{sec:introduction}

VTXO employs a Phased Fresnel Lens (PFL) to achieve near diffraction-limited angular resolution in the X-ray band\cite{krizmanic2020vtxo,rankin2020formation,pirayesh2018attitudea,pirayesh2019attitude,pirayesh2019deep,pirayesh2018attitudeb,pirayesh2017hybrid}. This lens offers the VTXO mission an imaging resolution of around 55 milli-arcsecond (mas) for a 1 km focal length in the X-ray spectrum. To maintain this level of accuracy, VTXO makes a precise alignment formation flying between its Inner Satellite (IS) and Outer Satellite (OS) with a sub-millimeter transverse accuracy. Formation flying in space provides advantages over a single spacecraft, including robustness, redundancy, reconfiguration, and broader coverage.

When the two spacecraft achieve this 1 km, 55 mas alignment precision, especially at their highest point, or apogee, they can begin their scientific observations. Notably, this 55 mas resolution is about ten times finer than the Chandra X-rays and the James Webb Space Telescope's 0.5 arcsecond resolutions\cite{krizmanic2020vtxo, meza2005line}. Such clarity from VTXO enables unprecedented studies of exoplanets\cite{pong2014high} and the environments surrounding space entities like black holes, neutron stars, and other stellar systems.

The Navy Interferometric Star Tracker Experiment II (NISTEx-II) \cite{Chester} plays a critical role in the precision formation flying of the VTXO mission by allowing highly accurate relative navigation. Mounted on the optical bench of (IS), the NISTEx-II star tracker, in conjunction with laser beacons installed on the Outer Satellite (OS), allows the GNC system to maintain sub-millimeter transverse alignment and attitude accuracy within 55 milli-arcseconds \cite{rankin2020formation,rankin2018vtxo,krizmanic2020vtxo}. To achieve this level of control, VTXO integrates thrusters, reaction wheels, an inertial measurement unit (IMU), GPS, the NISTEx-II star tracker, laser beacon systems, and a radio frequency (RF) ranging system that also serves as the inter-satellite communication link. These integrated sensing and actuation subsystems collectively establish stringent requirements on the Attitude Control System (ACS) state error \cite{krizmanic2020vtxo}. Additionally, energy consumption and mission duration are tightly constrained, driving the optimization of control strategies to minimize both, in line with mission-level requirements for efficiency and scientific yield.

The Navy Interferometric Star Tracker Experiment II (NISTEx-II) instrument \cite{Chester} provides accurate formation navigation for VTXO. Using NISTEx-II star tracker on the optical bench of IS and laser beacons on OS, GNC keeps the formation within millimeter level transverse alignment and 55 milli-arcsecond attitude accuracy\cite{rankin2020formation,rankin2018vtxo,krizmanic2020vtxo}. GNC of VTXO uses thrusters, reaction wheels, inertial navigation system sensor (IMU), GPS, NISTEx II precision star tracker on the optical bench of IS, laser beacons on the OS, and a radio ranging system that also serves as an inter-satellite communication link. The instrumentation of VTXO \cite{krizmanic2020vtxo} imposes requirements on the ACS state error. Beside, consumed energy of ACS and the time of the mission are desired to be minimal as the requirements of the VTXO.

The VTXO's relative position control system is divided into four stages, while its Attitude Control System (ACS) has three stages. Using trajectory optimization, we identify the best stages, orbits, and control strategies for the Guidance, Navigation, and Control (GNC) system\cite{krizmanic2020vtxo, rankin2020trajectory,rankin2020formation, calhoun2012covariance}. For VTXO's purpose, three stages are set for the ACS. Each stage aims to enhance VTXO's efficiency by improving accuracy and saving energy\cite{pirayesh2019deep,pirayesh2019attitude, pirayesh2017hybrid, pirayesh2018attitudeb, naseri2018formation, pirayesh2018attitudea}. Implementing these stages to the mission provides several advantages:

\begin{itemize}
\item Each stage has a specific goal tailored to the mission's objectives and the equipment in use.
\item Both the artificial intelligence (AI) system and the controllers for each stage are crafted to meet these specific goals.
\end{itemize}

To ensure optimal science observation, the VTXO mission employs specific accuracy metrics for its pre-observation phase, and uses an objective function to minimize energy consumption, which in turn enhances mission longevity, reduces costs, and ensures high-resolution pointing while maintaining stability across all starting conditions.
Sliding-mode control (SMC) can provide robustness and asymptotic stability in the presence of noise, disturbances with the disadvantages of chattering \cite{bai2018adaptive,markley2014fundamentals}. Chattering has to stay minimal in VTXO since NISTEx-II requires the angular velocities to stay close to zero. Other approaches including robust control \cite{li2007robust,luo2005h} including SMC, adaptive control\cite{wallsgrove2005globally, bai2018adaptive}, Deep neural networks (DNN) controllers for nonlinear system \cite{lewis2020neural} have shown asymptotic stability in the presence of disturbances and noise. However, usually adaptive controllers, including model reference adaptive control (MRAC), are developed for linear systems \cite{kodalak2015model, yechiel2017survey, lewis2020neural} and they are relatively slow in convergence. The nonlinear adaptive controller version developed by State Dependent Riccati Equations (SDRE) doesn't provide global asymptotic stability \cite{kodalak2015model}, as it only provides local asymptotic stability. DNN controllers, on the other hand, are stable globally and they can estimate the nonlinear dynamics \cite{lewis2020neural}, but, still the slow convergence is a concern, and the process of designing the controller is complex \cite{lewis2020neural}. In VTXO, the speed of controller convergence is important as it defines $T$ for the transient phase, and the dynamics of the nonlinear system is known. 
Multiplicative extended Kalman filter (MEKF) \cite{pirayesh2019attitude,pirayesh2018attitudeb, markley2014fundamentals,Woffinden2007,carpenter2018navigation} can provide the state estimation with noisy sensor measurements and having disturbances and uncertainties in the system. Extended state observer (ESO) \cite{bai2018adaptive} have shown promising solutions to the disturbance and uncertainty estimation and rejection while providing asymptotic stability. ESO estimates the time varying differentiable uncertainties and disturbances in the system, and the controller compensates for them in a single spacecraft \cite{bai2018adaptive} and multiple spacecraft formation control\cite{ye2017extended}.
Quaternions offer a robust and computationally efficient framework for solving equations of motion in spacecraft dynamics. Unlike Euler angles, they do not suffer from singularities and are straightforward to normalize \cite{markley2014fundamentals,sveier2019applied}. Quaternions are not exclusive to aerospace applications; they find use in fields as diverse as fluid mechanics and quantum mechanics \cite{kou2018linear}. 
In this paper, Monte-Carlo simulation is used for control system robustness and stability demonstration, and it is used to show control system robustness against external disturbances and noise measurement. Monte-Carlo simulation is used to show the control system robustness and stability for flexible robot arm \cite{ray1993monte} and for the formation control robustness of UAVs \cite{chaojie2014monte}. Sliding-mode control (SMC) and Lyapunov-based controller are used in different phases of the mission to ensure robustness and guarantee stability even when faced with external disturbances and noise. The Multiplicative Extended Kalman Filter (MEKF)\cite{pirayesh2019attitude,pirayesh2018attitudeb, markley2014fundamentals,Woffinden2007,carpenter2018navigation} estimates system states when there are disturbances, uncertainties, and noisy sensor data. Similar to the VTXO mission, a case study is done on the MASSIM X-ray virtual telescope with the astrometric sensor and the accelerometer sensors \cite{calhoun2012covariance}. The MASSIM mission is the precise formation between 2 spacecraft with 1000 kg mass and 1000 km distance. In the MASSIM mission case study, the 3-sigma estimation error of navigation structures is shown versus bias, noise, and disturbances. The controller is a linear controller only performed by the follower. It is shown that for different bias, noise, and disturbances, the navigation structure can provide different navigation estimation 3-sigma error.
This paper provides the VTXO mission with methods to increase the performance of VTXO by the following contributions
\begin{itemize}
    \item Test introducing timed automata method as a supervisory adaptive control for the analysis of phases of VTXO mission.
    \item Test using Monte-Carlo simulation to analyze the VTXO mission and the stability of the timed automata method.
\end{itemize}

While modern control strategies increasingly use AI, their application in safety-critical space missions is often hindered by their "black-box" nature. A lack of transparency in how an AI derives its control decisions creates significant challenges for verification, validation, and operator trust. This paper addresses this gap by proposing an Explainable AI (XAI)-enhanced framework. Our approach provides explainability on two levels: (1) At the AI level, the system does not merely output controller gains, but also predicts the resulting performance metrics (energy and error), making the AI's trade-offs transparent and interpretable. (2) At the supervisory level, a timed automata model provides a formal, deterministic, and verifiable logic for all mission phase transitions. This dual-level explainability ensures that the system's behavior is both optimal and understandable.

This paper provides the VTXO mission with methods to increase the performance of VTXO by the following contributions
\begin{itemize}
    \item Introducing timed automata method as a supervisory adaptive control for the analysis of VTXO mission.
    
    \item Introducing a new formation relative position control for VTXO

    \item Integrating a performance-predicting controller with a formally verifiable supervisory system to ensure trustworthy spacecraft autonomy.
    
    \item Introducing a neural network based solver for a unified constrained, non-convex dynamic optimization.

\end{itemize}

The codes and data in this article are available at https://github.com/Rpirayesh. 

\section{ VTXO Mission Concept of Operation and Its Implications for Supervisory Control}
A clear understanding of the Virtual Telescope for X-ray Observation (VTXO) mission's operational phases and accuracy requirements is fundamental to designing an effective supervisory control system. The intricate sequence of operations, each with distinct control objectives and constraints, dictates the need for an intelligent and adaptive supervisory layer capable of managing diverse control strategies. This section summarizes the concept of the VTXO mission, largely based on previous work \cite{pirayeshshirazinezhad2022designing, pirayesh2019attitude, pirayesh2019deep, pirayesh2018attitudeb}, to provide the necessary context for the development of XAI, the supervisory framework, and relative position control presented in this paper. The supervisory control system ensures seamless transitions between these phases, optimizes performance according to phase-specific metrics, and maintains stringent accuracy under varying conditions, all of which are directly informed by the mission profile detailed below.
In the VTXO mission, each orbit involves distinct operational phases for both the Attitude Control System (ACS) and the relative position formation between the Outer Satellite (OS) and the Inner Satellite (IS). These phases, first described in \cite{pirayeshshirazinezhad2022designing}, are critical for the subsequent control design.

\subsection{Phases of ACS}
ACS has 3 phases as follows:

\begin{itemize}
  \item The formation phase.
    \item The transient phase.
        \item The science phase.
    \end{itemize}

In the formation stabilization phase, the spacecraft are stabilized while they pass the perigee to come into the next orbit phase. GPS is used for orbit determination. The controller is a feed-forward controller that compensates for the external torque of the gravity gradient.

 During the transient phase, a Lyapunov controller is employed to reduce energy $E$ and guide convergence toward the equilibrium point $\pmb{x}_e$, for any initial conditions \cite{pirayeshshirazinezhad2022designing}. ACS aligns OS and IS for the observation phase. For navigation, Inertial Measurement Units (IMU) and star trackers are used. To maximize the duration of the observation and stabilization phases, the transient phase is constrained to a few minutes to increase the science observation period.
 
The science phase begins the precise formation after the transient phase with the following specifications \cite{pirayeshshirazinezhad2022designing, pirayesh2019deep,pirayesh2019attitude,pirayesh2018attitudeb,pirayesh2018attitudeb}.


\begin{itemize}
  \item The optical system must achieve an angular resolution of 55 milli-arcseconds (FWHM).
  \item A sub-millimeter level of accuracy is required for the transverse alignment between OS and IS.
  \item Pointing accuracy for each individual spacecraft should be within a few arcminutes.
  \item The distance between the OS and IS must be maintained at 1 km, with an allowable deviation of up to one meter.
\end{itemize}

This 55 milli-arcsecond angular resolution corresponds to maintaining sub-millimeter transverse alignment over the 1 km focal length; thus, the stringent alignment requirement, rather than the relative spacecraft orientation, drives the optical performance of the VTXO telescope.

For navigation, IS and (OS) are equipped with Inertial Measurement Units (IMU) and star trackers. The IS additionally uses laser beacons and radio ranging for enhanced accuracy. GPS is ruled out for high-altitude apogee navigation due to its limited accuracy. The Multiplicative Extended Kalman Filter (MEKF) and Sliding-Mode Control (SMC) are employed to ensure robustness against disturbances and model uncertainties\cite{pirayesh2019attitude}.

\subsection{Position Formation Phases}

Position formation is broken down into four main phases:

\begin{itemize}
\item De-formation phase
\item Tracking phase
\item Formation phase
\item Science phase
\end{itemize}

During the de-formation phase, the 1 km formation is broken and the satellites drift apart. The tracking phase uses GPS and radio ranging for orbit determination and kicks in near perigee. This phase is also when data, including images, is transmitted to Earth. The formation phase aims to get the IS and OS back to a distance of 1 km, meeting certain accuracy requirements. Finally, the science phase begins, where high-resolution imaging of space objects occurs.

\subsection{Accuracy Requirements for the transient phase}

Two primary accuracy requirements are imposed by the Field of View (FoV) of the NISTEx-II star tracker and VTXO. Percentile is used to characterize the accuracy requirements. $P_{99}(e)$ is used to characterize the accuracy requirements for the slew error.
Using $P_{99}$, 99 percent of times the mission meets the accuracy requirements.  

$R1$: \textbf{Accuracy requirements due to the formation of NISTEx-II instrument on IS and laser beacons on OS}

\begin{itemize}

  \item $R1_A$:$P_{99}(e)$ $< 5^\circ$
  \item $R1_B$:$P_{99}(e)$ $< 5^\circ$
\end{itemize}

$R2$: \textbf{Accuracy requirements due to the FoV of VTXO}

\begin{itemize}
  \item $R2_A$:$P_{99}(e)$  $< 0.18^\circ$
  \item  $R2_B$:$P_{99}(e)$ $< 0.18^\circ$
\end{itemize}

\subsection{Attitude representation and notation}
The estimated value or the value in the flight computer for the true variable $x$ is denoted as $\hat{x}$, and the measured value for $x$ is denoted as $\tilde{x}$. $\text{diag}(\pmb{x})$ represents the diagonal matrix with elements of vector $\pmb{x}$ in the diagonal.

The Earth coordinate frame is the Earth-centered inertia (ECI) frame,, and the frames used for the satellite body frame $F_b$ are the Local-Vertical-Local-Horizontal (LVLH) frames. 



The position of the spacecraft in the body frame is denoted as $\pmb{r}$.
Quaternions are used to model the attitude dynamics of equations\cite{markley2014fundamentals}.

Quaternion $\pmb{q}$ is a four-element vector with a three-element vector $\pmb{q}_{1:3}$ and a scalar part $q_4$. The quaternion unity constraint is 

\begin{equation} \label{QConstraint}
q_4^2=1-\pmb{\partial q}^T_{1:3}\pmb{\partial q}_{1:3}
\end{equation}

The orientation of frame $F_a$ to frame $F_b$ through Euler angle $\theta$ and Euler axis $\pmb{\upsilon }$ is expressed through quaternion as

\begin{equation} \label{RotationQ}
\pmb{q}^{ba}=\begin{bmatrix}
 \pmb{\upsilon }sin(\theta/2) \\\ cos(\theta/2)\\
\end{bmatrix}
\end{equation}

The associated attitude matrix corresponding to the quaternion $\pmb{q}^{ba}$ is shown as $\pmb{R}^{ba}$. 
 $\otimes$ is quaternion multiplication operator defined by \eqref{QOtimes}.

\begin{equation} \label{QOtimes}
\pmb{q} \otimes \bar{\pmb{q}}=\begin{bmatrix}
  q_4&-q_3&q_2\\q_3&q_4&-q_1\\-q_2&q_1&q_4&\\-q_1&-q_2&-q_3
\end{bmatrix}
\begin{bmatrix}
  \bar{q}_1\\\ \bar{q}_2\\\ \bar{q}_3 \\\ \bar{q}_4\\
\end{bmatrix}
\end{equation}

Quaternion multiplication between reference frames are 

\begin{equation} \label{MultiQ}
\pmb{q}^{ac}=\pmb{q}^{bc}\otimes\pmb{q}^{ab}
\end{equation}

Equation \eqref{MultiQ} corresponds to the multiplication of attitude matrices as
\begin{equation} \label{MultiQR}
\pmb{R}^{ac}=\pmb{R}^{bc}\pmb{R}^{ab}
\end{equation}

Small rotations $\theta$ can be written in terms of attitude matrix $\pmb{R}$ as

\begin{equation} \label{SmallRotation}
\partial \pmb{R}=\pmb{I}_{3\times3}-[\pmb{\theta}\times]
\end{equation}

Small rotations $\theta$ can also be written in terms of quaternion $\pmb{q}$ as

\begin{equation} \label{SmallRotationQ}
\partial \pmb{q}\approx\begin{bmatrix}
\pmb{\theta}/2 \\ 1
\end{bmatrix}
\end{equation}

Where $\pmb{\theta}=\theta\pmb{\upsilon }$, and $\pmb{I}$ is an identity matrix. The cross product matrix $[\pmb{x}\times]$ of the variable $\pmb{x}$ is defined as

\begin{equation} \label{CrossPr}
[\pmb{x}\times]=\begin{bmatrix}
  0 & -x_3 & x_2\\ x_3 & 0 & -x_1\\ -x_2 & x_1 & 0
\end{bmatrix}
\end{equation}


The identity quaternion is 
\begin{equation} \label{qI}
\pmb{q}=\begin{bmatrix}
  0 \\ 0 \\ 0 \\ 1
\end{bmatrix}
\end{equation}

\subsection{Orbits and the desired trajectories}

The baseline flight dynamics of VTXO uses a highly-elliptical geostationary transfer orbit with a 32.5-hour period for providing a 10-hour observation in the apogee \cite{rankin2025baseline}. The 5 Keplerian elements are the same for OS and IS except the eccentricity $\gamma$ given in table \ref{Keplerian elements}. 

The eccentricity of OS and IS are designed to include a few minutes of buffer between the time the OS and the IS pass the point where the orbits intersect, avoiding a collision between satellites. A larger difference between the eccentricities results in a lower risk of collision, since the satellites would have longer relative distances. However, this results in a higher energy consumption that is needed to keep the desired 1 km relative distance between the satellites.

The list of objects to be observed by VTXO are given in Table \ref{Objects}. This table of desired objects for VTXO can be updated in the future.

The desired attitude trajectory to observe the desired objects given in Table \ref{Objects} is denoted as $\pmb{q}_f$. $\pmb{q}_f$ doesn't vary by time; it just switches from one space object to the next when the science observation satisfies the time of observation.

\begin{table}[h!]
\caption{Time of observation for the objects VTXO observe in the observation phase \cite{krizmanic2020vtxo}.}
\label{Objects} 
\setlength{\tabcolsep}{3pt}
\begin{tabular}{|p{90pt}|p{150pt}|}
\hline
 Objects & 
Time of observation (hours)
\\
\hline
Sco X-1 & 0.2 \\\hline

GX 5-1 & 1.5 \\\hline

GRS 1915+105 & 4.2  \\\hline

Cyg X-3 & 4.9 \\\hline

Crab Pulsar & 5.4 \\\hline

Cen X-3 & 19 \\\hline

$\gamma$Cas & 146 \\\hline

Eta Carinae & 452 \\\hline

\end{tabular}
\end{table}

\subsection{ VTXO Dynamics}
$\pmb{q}^{bi}$ represents the orientation of the spacecraft body frame with respect to the earth Earth-centered inertial (ECI) frame, and $\pmb{\omega}_b^{bi}$ corresponds to slew rate of the spacecraft body frame with respect to the inertial frame.

The states $\pmb{x}$ for the attitude of spacecraft include quaternion and angular velocities, and the states $\pmb{s}$ for the position of spacecraft include the position and velocity of spacecraft. $\pmb{x}$ is given by

\begin{equation} \label{States}
\pmb{x}=\begin{bmatrix}
   \pmb{q}^T & \pmb{\omega}^T
\end{bmatrix}
^T
\end{equation}
\begin{equation*} \label{StatesQ}
\pmb{q}=\begin{bmatrix}
   q_1&q_2&q_3&q_4
\end{bmatrix}
^T
\end{equation*}
\begin{equation*} \label{StatesO}
\pmb{\omega}=\begin{bmatrix}
   \omega_1&\omega_2&\omega_3
\end{bmatrix}
^T
\end{equation*}

And $\pmb{s}$ is given by
\begin{equation} \label{rStates}
\pmb{s}=\begin{bmatrix}
   \pmb{r}^T &  \pmb{\text{v}}^T
\end{bmatrix}
^T
\end{equation}

$\pmb{x}$ at the beginning of the transient phase are the initial slew and initial slew rate for the transient phase denotes as $\pmb{x_0}$, and $\pmb{x}$ at the end of the transient phase are denotes as $\pmb{x_f}$.

A time variant modeling error $\delta \pmb{J}$ is considered in the nominal inertial momentum $\bar{\pmb{J}}$ to model the changes in the true value of inertial momentum $\pmb{J}$ as the following 

\begin{equation} \label{Mom}
\pmb{J}^{-1}=\bar{\pmb{J}}^{-1}+\delta \pmb{J}^{-1}
\end{equation}

The total applied torque to the spacecraft due to control input, noise, and disturbances are denoted as $\pmb{\tau}$.  The equations of motion for the dynamics of spacecraft are given by

\begin{equation} \label{Quaternions}
\dot{\pmb{q}}=\frac{1}{2}\pmb{\omega}\otimes\pmb{q}
\end{equation}

\begin{equation} \label{Omega}
\dot{\pmb{\omega}}=\pmb{J}^{-1}(\pmb{\tau}-\pmb{\omega}\times\pmb{J}\pmb{\omega})
\end{equation}

$\pmb{\tau}$ is given by


\begin{equation} \label{tauComp}
\pmb{\tau}=\pmb{\tau}_{in}+\pmb{{\tau}}_g+\pmb{w}_{\dot{\pmb{\omega}}}
\end{equation}
$\pmb{\tau}_{in}$ corresponds to the control input, $\pmb{w}_{\dot{\pmb{\omega}}}$ corresponds to external disturbances from the space environment modeled as Gaussian white noise, and $\pmb{\tau}_g$ corresponds to the time variant gravity gradient torque. $\pmb{\tau}_{in}$ derives the quaternion \pmb{q} to the desired quaternion $\pmb{q}_f$ and $\pmb{\omega}$ to zero, which is equal to reaching the equilibrium point $\pmb{x}_e$. 



The dynamics for the spacecraft relative position formation is given by

\begin{equation} \label{Velocityr}
\dot{\pmb{r}}=\pmb{\text{v}}
\end{equation}

\begin{equation} \label{rDoubleDot}
\ddot{\pmb{r}}=- \sum_{i=1}^{n} \mu_i \frac{\pmb{r}_i}{|\pmb{r}_i|^3}+\pmb{u}
\end{equation}

$\pmb{r}_{i}$ shows the position vector of the spacecraft to the gravitational bodies.

The relative position dynamics is given by

\begin{equation} \label{Relative}
\pmb{r}_{rel}=\pmb{r}_{F}-\pmb{r}_{L}
\end{equation}

$\pmb{u}$ is given by

\begin{equation} \label{u}
\pmb{u}=\pmb{u}_{g}+\pmb{u}_{in}
\end{equation}

$\pmb{u}_{in}$ corresponds to the position controller provided by the thrusters, and $\pmb{u}_{g}$ corresponds to solar disturbances $\pmb{g}_{\text{solar}}$ and other perturbations $\pmb{g}_{\text{pert}}$ given by

\begin{equation} \label{ug}
\pmb{u}_{g}=\pmb{g}_{\text{solar}}+\pmb{g}_{\text{pert}}
\end{equation}

As a result, the relative position dynamics is given by

\begin{equation} \label{Relativedot}
\dot{\pmb{r}}_{rel}=\pmb{\text{v}}_F-\pmb{\text{v}}_L
\end{equation}

\begin{equation} \label{RelativeDdot}
\ddot{\pmb{r}}_{rel}=- \sum_{i=1}^{n} \mu_i (\frac{\pmb{r}_{iF}}{|\pmb{r}_{iF}|^3}-\frac{\pmb{r}_{iL}}{|\pmb{r}_{iL}|^3})+\pmb{u}_F-\pmb{u}_L
\end{equation}

$\pmb{r}_{iF}$ and $\pmb{r}_{iL}$ show the vector of the follower and leader to the gravitational bodies, respectively.
\subsection{External disturbances and gravity gradient torque} \label{Dist}

In VTXO, disturbances are modeled as bounded, time-varying, differentiable forces acting on the position and torques in the dynamics of the spacecraft.

Gravity gradient torque $\pmb{\tau}_g$ is derived from point mass gravity models \cite{markley2014fundamentals,woffinden2008angles} as the following

\begin{equation} \label{ToyG}
\pmb{\tau}_g=\frac{3\mu}{|\pmb{r}|^3}\pmb{n}\times(\pmb{J}\pmb{n})
\end{equation}
where \pmb{n} is the nadir vector in the body frame $F_b$, and $|\pmb{r}|$ is the radial distance of spacecraft from the earth.

 $\pmb{w}_{\dot{\pmb{\omega}}}$ corresponds to the random torques, J2 gravity model, and torques to account for drag, solar pressure, higher-order-gravity terms, etc. $\pmb{w}_{\dot{\pmb{\omega}}}$ is modeled as a zero-mean Gaussian white noise process where the power of the noise is captured in the variance as $\sigma^2_{\dot{\pmb{\omega}}}$\cite{lear1985kalman}.

\begin{equation} \label{Toyd}
\mathbb{E}[\pmb{w}_{\dot{\pmb{\omega}}}(t)\pmb{w}_{\dot{\pmb{\omega}}}(t')^T]=\sigma^2_{\dot{\pmb{\omega}}}\pmb{I}_{3\times3}\delta(t-t')
\end{equation}

$\pmb{u}_{g}$ is modeled as a first-order Markov processes also known as exponentially correlated random variables (ECRV) given by

\begin{equation} \label{ugModel}
\dot{\pmb{u}}_{g}=\frac{-\pmb{u}_{g}}{\tau_g}+\pmb{w}_g
\end{equation}

Large and small time constant $\tau$ makes the bias a constant value or a white noise, respectively.

The variance of Gaussian white noise $\pmb{w}_g$ is given by $\sigma^2_g$, and the time constant for $\pmb{u}_g$ is given by $\tau_g$.

$\delta(t-t')$ is the Dirac delta function defined as 

\begin{equation} \label{DeltaDirac}
\delta(t-t')=0 \quad \text{if} \quad t\neq t' 
\end{equation}

\begin{equation} \label{DeltaDiracInteg}
\int_{-\infty}^{\infty} \delta(t-t') \,dt'=1
\end{equation}

$\sigma^2_{\dot{\pmb{\omega}}}$ are chosen based on the orbit and the size of spacecraft. If the spacecraft are bigger or closer to earth, $\sigma^2_{\dot{\pmb{\omega}}}$ is chosen larger due to higher drag forces and other torques. Choosing the external disturbances on the spacecraft is shown by Lear\cite{lear1985kalman,Woffinden2007}, where the estimated disturbance corresponds to the expected downrange and attitude error after one orbit. In \cite{calhoun2012covariance}, the external disturbances are modeled as first-order Markov processes while in \cite{Woffinden2007} it is given as a white noise.
\subsection{Sensor model VTXO}

Gyroscope in inertial measurement unit (IMU) is used to measure $\pmb{\omega}$ given by $\tilde{\pmb{\omega}}$,  star tracker is used for measuring $\pmb{q}$ given by $\tilde{\pmb{q}}$, accelerometer in IMU for measuring $\tilde{\ddot{\pmb{r}}}$, Radio ranging sensor for measuring $\tilde{r}^z_{rel}$, and interferometry sensor in the NISTEx-II star tracker for measuring $\tilde{\pmb{r}}^{xy}_{rel}$.

\subsection{Actuator model VTXO}

In VTXO, reaction wheels are used for ACS and thrusters are used for relative position control as the primary actuators in GNC for both the follower and the leader. The torque generated by the reaction wheels $\pmb{\tau}_{in}$ and by the thrusters $\pmb{u}_{in}$ use the commanded torque $\hat{\pmb{\tau}}_{in}$ and commanded force $\hat{\pmb{u}}_{in}$, respectively, given by the control law.

\subsubsection{Reaction wheel}

In $\pmb{\tau}_{in}$ \cite{Woffinden2007,mok2020performance}, Gaussian white noise $\pmb{w}_\tau$, bias $\pmb{b}_\tau$, scale factor $\pmb{f}_\tau$, and misalignment $\pmb{\epsilon}_\tau$ are included as the following

\begin{equation} \label{ToyRW}
\pmb{\tau}_{in}=\partial \pmb{R}(\pmb{\epsilon}_\tau)[\{\pmb{I}_{3\times3}+\text{diag}(\pmb{f}_\tau)\}\hat{\pmb{\tau}}_{in}+\pmb{b}_\tau+\pmb{w}_\tau]
\end{equation}

The variance $\sigma^2_{\pmb{w}_\tau}$ captures the power of the noise in the random noise $\pmb{w}_\tau$ as

\begin{equation} \label{SigmaRW}
\mathbb{E}[\pmb{w}_\tau(t)\pmb{w}_\tau(t')^T]=\sigma^2_{\pmb{w}_\tau}\pmb{I}_{3\times3}\delta(t-t')
\end{equation}

The bias $\pmb{b}_\tau$ is modeled as a first-order Markov processes given by

\begin{equation} \label{toy1aBias}
\dot{\pmb{b}}_\tau=\frac{-\pmb{b}_\tau}{\tau_\tau}+\pmb{w}_{b \tau}
\end{equation}

The variance $\sigma^2_{\pmb{w}_{b \tau}}$ captures the power of the noise in the random noise $\pmb{w}_{b \tau}$.

\subsubsection{Thruster}

In $\pmb{u}_{in}$ \cite{Woffinden2007,mok2020performance, calhoun2012covariance}, Gaussian white noise $\pmb{w}_u$, bias $\pmb{b}_u$, scale factor $\pmb{f}_u$ are included in the model as the following

\begin{equation} \label{uThruster}
\pmb{u}_{in}=[\{\pmb{I}_{3\times3}+\text{diag}(\pmb{f}_u)\}\hat{\pmb{u}}_{in}+\pmb{b}_u+\pmb{w}_u]
\end{equation}

The variance $\sigma^2_{\pmb{w}_u}$ captures the power of the noise in the random noise $\pmb{w}_u$ as

\begin{equation} \label{SigmaUu}
\mathbb{E}[\pmb{w}_u(t)\pmb{w}_u(t')^T]=\sigma^2_{\pmb{w}_u}\pmb{I}_{3\times3}\delta(t-t')
\end{equation}

The bias $\pmb{b}_u$ is modeled as a first-order Markov processes given by

\begin{equation} \label{OmegaBiasU}
\dot{\pmb{b}}_u=\frac{-\pmb{b}_u}{\tau_u}+\pmb{w}_{b u}
\end{equation}

The variance $\sigma^2_{\pmb{w}_{b u}}$ captures the power of the noise in the random noise $\pmb{w}_{b u}$.

\subsection{Navigation structure and sensor fusion}\label{NV}
The estimated value or the value in the flight computer for the true variable $x$ is denoted as $\hat{x}$, and the measured value for $x$ is denoted as $\tilde{x}$.

The variation of EKF with the model replacement method and the multiplicative extended Kalman filter (MEKF) are used in VTXO. In the transient phase and the science observation phase, without keeping the precision formation, star trackers and gyro sensors are used to provide a few mas accuracy for both follower and leader.

In the science observation phase, 55 mas formation accuracy, sub millimeter formation transverse alignment accuracy, and 1 meter relative distance accuracy are obtained using the following instruments

\begin{itemize}
  \item Gyro sensor.
  \item  Accelerometer sensor.
    \item Radio ranging.
    \item Laser beacons on the leader spacecraft.
        \item NISTEx-II star tracker on the follower spacecraft.
        \item Star tracker on the leader.
    \end{itemize}
    
In the science observation phase, both attitude control and relative position control are used to hold the precise formation.

For the relative position in the phases before and after the science observation phase, the accelerometer sensor and the GPS are used. In the formation stabilization phase, GPS is used. GPS calibrates the orbits to correct any deviation from the nominal orbits at the perigee.  

 \subsubsection{Dynamics of spacecraft for prediction in the navigation states}

The spacecraft dynamics used for propagating the states are as follows. 

\begin{equation} \label{HatQuaternions}
\dot{\hat{\pmb{q}}}=\frac{1}{2}\hat{\pmb{\omega}}\otimes\hat{\pmb{q}}
\end{equation}

\begin{equation} \label{HatOmega}
\hat{\dot{\pmb{\omega}}}=\hat{\pmb{J}}^{-1}(\hat{\pmb{\tau}}-\hat{\pmb{\omega}}\times\hat{\pmb{J}}\hat{\pmb{\omega}})
\end{equation}

\begin{equation} \label{HatR}
\dot{\hat{\pmb{r}}}=\hat{\pmb{\text{v}}}
\end{equation}

\begin{equation} \label{HatRd}
\dot{\hat{\pmb{\text{v}}}}=-\mu \frac{\hat{\pmb{r}}}{|\hat{\pmb{r}}|^3}-\hat{\pmb{u}}
\end{equation}

For the relative position, the propagated states are given as

\begin{equation} \label{RelativeP}
\hat{\pmb{r}}_{rel}=\hat{\pmb{r}}_{F}-\hat{\pmb{r}}_{L}
\end{equation}

\begin{equation} \label{RelativedotDotR}
\dot{\hat{\pmb{r}}}_{rel}=\pmb{\hat{\text{v}}}_F-\pmb{\hat{\text{v}}}_L
\end{equation}

\begin{equation} \label{HatRdRR}
\dot{\hat{\pmb{\text{v}}}}_{rel}=-\mu (\frac{\hat{\pmb{r}}_F}{|\hat{\pmb{r}}_F|^3}-\frac{\hat{\pmb{r}}_L}{|\hat{\pmb{r}}_L|^3})+\hat{\pmb{u}}_{F}-\hat{\pmb{u}}_{L}
\end{equation}

The first-order Markov process bias and misalignment are propagated as

\begin{equation} \label{OmegaBiasHatG}
\hat{\dot{\pmb{b}}}=\frac{-\pmb{b}}{\tau}
\end{equation}

\begin{equation} \label{MisBiasHatG}
\hat{\dot{\pmb{\epsilon}}}=\frac{-\pmb{\epsilon}}{\tau}
\end{equation}

\subsubsection{Multiplicative extended Kalman filter for the attitude control system}

The multiplicative extended Kalman filter (MEKF) with model replacement mode is used for the ACS.

When the model replacement mode is used, the angular velocity is directly replaced with the gyro measurements $\tilde{\pmb{\omega}}$, and the spacecraft dynamics is propagated as 
\begin{equation} \label{HatQuaternionsPro}
\dot{\hat{\pmb{q}}}=\frac{1}{2}(\tilde{\pmb{\omega}}-\hat{\pmb{b}}_{\omega})\otimes\hat{\pmb{q}}
\end{equation}

Since the measurements violate the quaternion unity constraint \eqref{QConstraint}, the multiplicative version of EKF is used to propagate the quaternion states. In MEKF , the propagated quaternions are updated by small orientation \eqref{SmallRotationQ} given by
\begin{equation} \label{MEKFtheta}
\pmb{q}^+=\pmb{\partial{q}}^+(\pmb{\theta}) \otimes \pmb{q}^-
\end{equation}

When MEKF is used, the navigation states of each satellite is a 12-element vector given by
\begin{equation} \label{StatesNavigation12}
\pmb{x}=\begin{bmatrix}
  \pmb{\theta}_b & \pmb{\omega} & \pmb{b}_{\omega} & \pmb{b}_{q}
\end{bmatrix}
^T
\end{equation}

When MEKF with model replacement is used, the navigation states of each satellite is a 9-element vector given by
\begin{equation} \label{StatesNavigation9}
\pmb{x}=\begin{bmatrix}
  \pmb{\theta}_b & \pmb{b}_{\omega} & \pmb{b}_{q}
\end{bmatrix}
^T
\end{equation}

Further discussion on MEKF with model replacement for VTXO ACS is given in \cite{pirayesh2019attitude}.

\subsection{Control law}\label{Control}

A nonlinear globally asymptotic stability Lyapunov controller \cite{markley2014fundamentals,wie1985quaternion} is defined as the control law for the transient phase attitude control. SMC is defined for the science observation attitude control and relative position control. Anti gravity gradient torque is considered for the formation stabilization phase.

For showing the effect of linear controllers and the effect of SMC in the transient phase, they are implemented and compared with the chosen controllers.

For the ACS, the control law minimizes the difference between the desired quaternion $\pmb{q}_f$ and the current state of quaternions $\pmb{q}$ and the difference between the desired angular velocities $\pmb{w}_f$ and the current state of angular velocities  $\pmb{w}$. The difference between $\pmb{q}_f$ and $\pmb{q}$ is denoted by $\pmb{\partial q}$ given by
\begin{equation} \label{RoQ}
\pmb{\partial q}\equiv
\begin{bmatrix}
   \pmb{\partial q}_{1:3}\\ \partial q_4
\end{bmatrix}
=\pmb{q}\otimes {\pmb{q}_f}^{-1}
\end{equation}

\begin{equation} \label{EQ}
\pmb{\partial q}_{1:3}=\pmb{E}(\pmb{q}_f)^T\pmb{q}
\end{equation}

where $\pmb{E}(\pmb{q})$ is the 4$\times$3 matrix

\begin{equation} \label{EQ1}
\pmb{E}(\pmb{q})=\begin{bmatrix}
  q_4&-q_3&q_2\\q_3&q_4&-q_1\\-q_2&q_1&q_4&\\-q_1&-q_2&-q_3
\end{bmatrix}
\end{equation}

\begin{equation} \label{EQ2}
\partial q_4=\pmb{q}^T\pmb{q}_f
\end{equation}

Taking the time derivative of \eqref{RoQ} leads to

\begin{equation} \label{RoQDot}
\pmb{\partial \dot{q}}
=\pmb{\dot{q}}\otimes {\pmb{q}_f}^{-1}
\end{equation}

Using \eqref{RoQ} and substituting \eqref{Quaternions} into \eqref{RoQDot} gives

\begin{equation} \label{QdotPartials}
\pmb{\partial \dot{q}}_{1:3}=\frac{1}{2}[\pmb{\partial q}_{1:3}\times]\pmb{\omega}+\frac{1}{2}\partial q_4\pmb{\omega}
\end{equation}

\begin{equation} \label{Q4dotPartials}
\partial \dot{q}_4=-\frac{1}{2}\pmb{\partial q}_{1:3}^T\pmb{\omega}
\end{equation}

\subsubsection{Linear controller}\label{Lin}
For the attitude controller, a linear proportional-derivative (PD) controller is used as the following 
\begin{equation} \label{tau_inPD}
\hat{\pmb{\tau}}_{in}=\hat{k}^q_P\hat{\pmb{\partial q}_{1:3}}-\hat{k}^q_D(\hat{\pmb{\omega}}_f-{\pmb{\omega}})
\end{equation}

Where $\pmb{\omega}_f$ shows the final desired angular velocity which is zero in VTXO. The vector of the PD controller parameters to be defined for ACS is given by
 
 \begin{equation} \label{PDtoyACS}
\hat{\pmb{k}}=\begin{bmatrix}
   \hat{k}^q_P \\ \hat{k}^q_D 
\end{bmatrix}
\end{equation}

For the relative position controller, a linear proportional-derivative (PD) controller is used as the following 
\begin{equation} \label{PDrR}
\hat{\pmb{u}}_{in}=\hat{k}^r_P({\pmb{r}}^f_{rel}-\hat{\pmb{r}}_{rel})+\hat{k}^r_D({\pmb{\text{v}}}^f_{rel}-\hat{\pmb{\text{v}}}_{rel})
\end{equation}

 



\subsubsection{A Lyapunov controller for the attitude control system}
The Lyapunov controller is defined as the following
\begin{equation} \label{tau_in}
\hat{\pmb{\tau}}_{in}=-\hat{k_1}\text{sign}(\partial \hat{q_4})\hat{\pmb{\partial q}_{1:3}}-\hat{k_2}(1 - \hat{\pmb{\partial q}_{1:3}}^T\hat{\pmb{\partial q}_{1:3}})\hat{\pmb{\omega}}
\end{equation}

where $\hat{k}$ is estimated by Monte-Carlo, optimization, ML, or chosen randomly with a uniform distribution and $\hat{\pmb{x}}$, as explained earlier, is obtained from navigation filters. In \cite{pirayeshshirazinezhad2022designing, pirayeshshirazinezhad2022artificial}, the global asymptotic stability is proved using the Lyapunov stability theorem. 







In the simulation, noise, disturbances, and uncertainty are considered, and the stability for the closed-loop system is shown using Monte Carlo simulation.

\subsubsection{Sliding mode controller for the attitude control system} \label{SMCACS}
The sliding vector $\pmb{\text{s}}$ for ACS is given by \cite{markley2014fundamentals}

\begin{equation} \label{sliding vectorACS}
\hat{\pmb{s}}=(\hat{\pmb{\omega}}-{\pmb{\omega}}_f)+\hat{k}_{SMC}\text{sign}(\partial \hat{q_4})\hat{\pmb{\partial q}_{1:3}}
\end{equation}

As explained in section \ref{SMCL}, the control law is obtained by $\dot{\pmb{s}}=0$ and including the sat function. The control law is obtained as  

\begin{align} 
\begin{split}
\hat{\pmb{\tau}}_{in} = & \pmb{J}\Bigg\{\frac{\hat{k}_{SMC}}{2}\bigg[|\partial \hat{q_4}|({\pmb{\omega}}_f-\hat{\pmb{\omega}}) \\
& -\text{sign}(\partial \hat{q_4})\hat{\pmb{\partial q}_{1:3}} \times ({\pmb{\omega}}_f+\hat{\pmb{\omega}})\bigg] \\
& +\dot{\pmb{\omega}}_f- \hat{\pmb{Z}}\text{sat}(\hat{s_i},\hat{\varepsilon}_i)\Bigg\} \\
& +\hat{\pmb{\omega}}\times\pmb{J}\hat{\pmb{\omega}}
\end{split}
\label{slidingModeU}
\end{align}

$s_i$ shows the ith component of the sliding vector in \eqref{sliding vectorACS}, where $i=1,2,3$. $\pmb{Z}$ is a positive definite matrix, and $\varepsilon_i$ is a positive scalar. To reduce the chattering, instead of a sign function, the saturation function $\text{sat}(\hat{s_i},\hat{\varepsilon}_i)$ is used given as the following
\begin{equation} \label{SatF}
\text{sat}(\hat{s_i},\hat{\varepsilon}_i)=\begin{bmatrix}
  1 \quad \text{for} \quad  \hat{s_i} > \hat{\varepsilon}_i \\ \hat{s_i}/\hat{\varepsilon}_i  \quad \text{for} \quad  |\hat{s_i}| \leq \hat{\varepsilon}_i \\ -1  \quad \text{for} \quad  \hat{s_i} < \hat{\varepsilon}_i 
\end{bmatrix}
\end{equation}

The saturation function $\text{sat}(\hat{s_i},\hat{\varepsilon}_i)$ derives the system to the sliding surface. The following Lyapunov function proves the global asymptotic stability.

\begin{equation} \label{vsHS}
V(\Delta(\pmb{x}))=\frac{1}{2}\pmb{s}^2
\end{equation}

And the time derivative of the Lyapunov function is given as
\begin{equation} \label{vsdHS}
\dot V(\Delta(\pmb{x}))=\pmb{s}\dot{\pmb{s}}
\end{equation}

In section \ref{SMCL}, the proof for the global asymptotic stability of SMC is shown. Since the desired $\pmb{\omega}$ are zero, $\pmb{\omega}_f=0$. $\hat{k}$,$\hat{\varepsilon}_i$, and $\hat{\pmb{Z}}$ can be estimated by Monte-Carlo, optimization, ML, or chosen randomly with a given distribution. $\hat{\pmb{x}}$ is obtained from navigation filters given in section \ref{NV}.

 $\pmb{Z}$ is chosen to be a positive scalar here denoted as $z$. The vector of SMC parameters to be estimated is given by
 
 \begin{equation} \label{SMCtoy}
\hat{\pmb{k}}=\begin{bmatrix}
   \hat{z} \\ \hat{k}_{SMC} \\ \hat{\varepsilon} 
\end{bmatrix}
\end{equation}

\subsubsection{Sliding mode controller for the relative position control law} \label{SMCR}
In the MASSIM virtual telescope \cite{calhoun2012covariance}, it is assumed that the leader drifts in its natural orbit and the follower keeps the desired relative position with the leader. In the science observation phase, both the follower and the leader control the relative position and consider the thrust control input of the other spacecraft as a disturbance in the SMC. 
The sliding vector $\pmb{\text{s}}$ for the relative position control is given by

\begin{equation} \label{sliding vectorR}
\hat{\pmb{s}}=(\hat{\pmb{\text{v}}}_{rel}-{\pmb{\text{v}}}^f_{rel})+\hat{k}(\hat{\pmb{r}}_{rel}-{\pmb{r}}^f_{rel})
\end{equation}

When the relative position is controlled by the follower, the assumed model $\bar{f}(\hat{{\pmb{r}}}_{rel},\hat{\pmb{\text{v}}}_{rel}$) is given by

\begin{equation} \label{AModel}
\bar{f}(\hat{{\pmb{r}}}_{rel},\hat{\pmb{\text{v}}}_{rel})=-\mu (\frac{\hat{\pmb{r}}_F}{|\hat{\pmb{r}}_F|^3}-\frac{\hat{\pmb{r}}_L}{|\hat{\pmb{r}}_L|^3})-\hat{\pmb{u}}_{inL}
\end{equation}

For a natural drift of the leader, $\hat{\pmb{u}}_{inL}$ is zero. The ${\pmb{\text{v}}}^f_{rel}=0_{3\times1} $ and ${\pmb{\dot{\text{{v}}}}}^f_{rel}=0_{3\times1}$. 

As explained in section \ref{SMCL}, the control law is obtained by $\dot{\pmb{s}}=0$ and including the sat function. The control law is obtained as  

\begin{equation} \label{slidingModeRelU}
\hat{\pmb{u}}_{inF}=\mu (\frac{\hat{\pmb{r}}_F}{|\hat{\pmb{r}}_F|^3}-\frac{\hat{\pmb{r}}_L}{|\hat{\pmb{r}}_L|^3})+\hat{\pmb{u}}_{inL} -\hat{k} (\hat{\pmb{\text{v}}}_{rel}-{\pmb{\text{v}}}^f_{rel})-\hat{\pmb{Z}}\text{sat}(\hat{s_i},\hat{\varepsilon}_i)
\end{equation}

$s_i$ shows the ith component of the sliding vector in \eqref{sliding vectorR}, where $i=1,2,3$. $\hat{\pmb{Z}}$ is a diagonal positive definite matrix, and $\varepsilon_i$ is a positive scalar. To reduce the chattering, instead of a sign function, the saturation function $\text{sat}(\hat{s_i},\hat{\varepsilon}_i)$ is used given as the following
\begin{equation} \label{SatFF}
\text{sat}(\hat{s_i},\hat{\varepsilon}_i)=\begin{bmatrix}
  1 \quad \text{for} \quad  \hat{s_i} > \hat{\varepsilon}_i \\ \hat{s_i}/\hat{\varepsilon}_i  \quad \text{for} \quad  |\hat{s_i}| \leq \hat{\varepsilon}_i \\ -1  \quad \text{for} \quad  \hat{s_i} < \hat{\varepsilon}_i 
\end{bmatrix}
\end{equation}

The saturation function $\text{sat}(\hat{s_i},\hat{\varepsilon}_i)$ derives the system to the sliding surface. The following Lyapunov function proves the global asymptotic stability 

\begin{equation} \label{vsH}
V(\Delta(\pmb{x}))=\frac{1}{2}\pmb{s}^2
\end{equation}

And the time derivative of the Lyapunov function is given as
\begin{equation} \label{vsdH}
\dot V(\Delta(\pmb{x}))=\pmb{s}\dot{\pmb{s}}
\end{equation}

In section \ref{SMCL}, the proof of the global asymptotic stability of SMC is shown. $\hat{k}$,$\hat{\varepsilon}_i$, and $\hat{\pmb{Z}}$ can be estimated by Monte-Carlo, optimization, ML, or chosen randomly with a given distribution. $\hat{\pmb{x}}$ is obtained from navigation filters given in section \ref{NV}. With the same approach, the thruster input for the leader $\hat{\pmb{u}}_{inL}$, when the follower control input is considered disturbance, is obtained as 

\begin{equation} \label{slidingModeRelUL}
\hat{\pmb{u}}_{inL}=\mu (\frac{\hat{\pmb{r}}_L}{|\hat{\pmb{r}}_L|^3}-\frac{\hat{\pmb{r}}_F}{|\hat{\pmb{r}}_F|^3})-\hat{\pmb{u}}_{inF} -\hat{k} (\hat{\pmb{\text{v}}}_{rel}-{\pmb{\text{v}}}^f_{rel})-\hat{\pmb{Z}}\text{sat}(\hat{s_i},\hat{\varepsilon}_i)
\end{equation}

Where $\pmb{r}_{rel}$ and $\dot{\pmb{r}}_{rel}$ are defined as $\pmb{r}_{rel}=\pmb{r}_{L}-\pmb{r}_{F}$ and   $\dot{\pmb{r}}_{rel}=\pmb{\text{v}}_L-\pmb{\text{v}}_F$ , respectively.

\section{Artificial Intelligence Framework for Adaptive Attitude Control} 

ML estimates the optimal parameters in the transient and science phases using supervised learning. The data is obtained using optimization methods. The optimal parameters that ML estimates are $\pmb{k} = [k_1, k_2$], $E$, and $e$ based on $\pmb{x_f}$, $\pmb{x_0}$ and $w$. $T$ is estimated by ML for the transient phase, and it is chosen from the table \ref{Objects}.

\begin{figure}[ht!]
\centering
\includegraphics[width=0.9\linewidth]{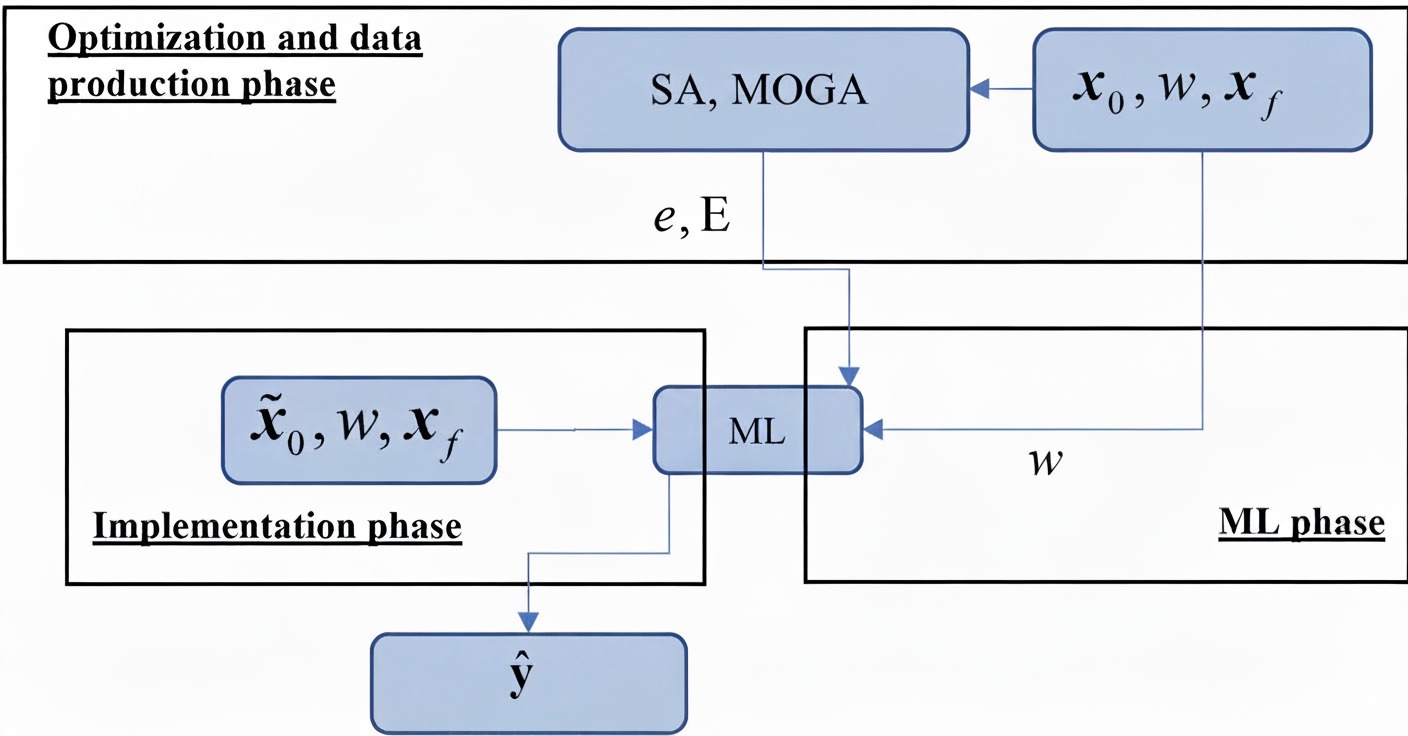}
\caption{This unified framework addresses both transient and science phases. In the optimization and data production phase, initial states ($\pmb{x_0}$), weights ($w$), and final states ($\pmb{x_f}$) are generated and given to optimization. Simulated Annealing (SA) and Multi-objective Genetic Algorithm (MOGA) optimization algorithms then optimize variables $k_1, k_2$, and transient phase duration $T$ to minimize the objective function $\chi$ (composed of $e$ and $E$). The resulting optimized dataset trains the Machine Learning (ML) model to predict the optimal parameters ($\pmb{\hat{y}}$)}
\label{fig2}
\end{figure}

Fig. \ref{fig2} shows the 3 stages of AI as follows:

\begin{itemize}
  \item Optimization and data production phase
    \item ML phase
    \item Implementation phase 
\end{itemize}

To produce each data, the objective function $\chi$ is optimized with the variable $\pmb{k}$ and $T$ (for the transient phase) for each set of $\pmb{x_0}$, $w$, and $\pmb{x_f}$. $\chi$ is a vector \eqref{UnifiedFormulation} of \text{$E$} and \text{$e$} for the science phase weighted sum of \text{$E$} and \text{$e$} \eqref{OF} for the transient phase. After producing the data, ML estimates $\pmb{y}=[\text{$E$},\quad \text{$e$},\quad k_1,\quad k_2]$ with the input $w$, $\pmb{x_0}$, and $\pmb{x_f}$. Also $T$ is estmated with ML in the tansient phase.

In the ML phase, the inputs to ML are $w$ and $\pmb{x_0}$, and the outputs of the ML are the \text{$E$}, $e$, and $\pmb{z}$ defined as $\pmb{y}=[\text{$E$},\quad \text{$e$},\quad k_1,\quad k_2]$.

The optimization and data production phase and ML phase are implemented offline, whereas the implantation phase is implemented in the real-time in space using FPGAs. In the implantation phase, the values of $\pmb{x_0}$ obtained from attitude estimation algorithms \cite{markley2014fundamentals, pirayesh2019attitude}, denoted by $\pmb{\tilde{x}}_0$, and the value of $w$ gives the estimated optimal $\pmb{k}$ denoted by $\pmb{\hat{k}}$. The ML estimates \text{$E$} and $e$, denoted by $\hat{E}$ and $\hat{e}$ respectively, obtained from the maneuver by the $\pmb{\hat{z}}$ parameters. The estimated $\pmb{y}$ used in the implementation phase is denoted by $\pmb{\hat{y}}$. It is shown later that DNN is chosen to estimate $\pmb{\hat{y}}$. The DNN is incorporated into FPGA, as the input to the FPGA is $\pmb{\tilde{x}}_0$, $w$ and the output of FPGA is $\pmb{\hat{y}}$.

The optimization and ML phases are conducted offline for training and testing. The implementation phase operates in real-time on FPGAs for inference, using attitude estimation algorithms \cite{markley2014fundamentals, pirayesh2019attitude} to provide $\pmb{\tilde{x}}_0$. With $w$ as an additional input, the FPGA-embedded DNN directly estimates the optimal parameters $\pmb{\hat{y}}$ for the optimal maneuver.

A key contribution of the artificial intelligence (AI) framework presented here is its inherent explainability, which is critical for mission-critical systems. Unlike opaque "black-box" models that simply provide an output, this framework offers transparency. The explainability is achieved as follows:

\begin{description}
    \item[\textbf{Predictive Transparency:}] The ML model is trained not only to determine the optimal controller parameters ($z$) but also to predict the associated energy consumption ($E$) and slew error ($e$). As shown in Figure~4, the output $\hat{y}$ includes these performance metrics. This allows a mission operator (or an autonomous system) to understand the consequences of a control decision before it is executed. For example, the system can explicitly manage the trade-off between performance and resource usage by adjusting the weight $w$, knowing the predicted outcome. This contrasts sharply with methods like adaptive LQR, where the future energy and error costs are not explicitly estimated as part of the real-time control loop.

    \item[\textbf{Structural Transparency:}] The overall mission logic is governed by the supervisory control system (detailed in Section~12), which uses a clear, rule-based timed automata. This ensures that every phase transition is deterministic and based on verifiable conditions, not on an uninterpretable internal state of a complex model.
\end{description}

The goal of this explainable AI is to provide ACS with optimal adaptive values, including the duration of the transient phase $T$.
\subsection{Unified Supervisory Optimization Formulation}


The optimization loss function $\chi$ for the transient phase is a scalar given as:

\begin{equation} \label{OF}
\chi= E+e
\end{equation}
The loss function $\chi$ of the Pareto optimization for the science phase is a vector given as:

\begin{equation} \label{OF}
\chi= \begin{bmatrix}
    E\\ e
\end{bmatrix} 
\end{equation}

$w=e/E$ is a weight for \text{$e$} over \text{$E$}. The scaler value of \text{$e$} is minimized for the whole duration of the science phase, while it is minimized in the last $D=4$ seconds \eqref{RMSE} to minimize the steady state tracking error in the transient phase. $D$ is chosen to be minimal to reduce the computation cost for the large angle manouvers, which also corresponds to the steady-state error.  
Variable \text{$e$} is defined as

\begin{equation} \label{RMSE}
\text{$e$}=\frac{1}{D}\int_{T_{end}-D}^{T_{end}} \sqrt{\pmb{e}(t)^\top\pmb{e}(t)} \, \,dt
\end{equation}
where $\pmb{e}(t)$ is the absolute difference between the desired attitude and the actual attitude. $T_{end}$ is the end of the phase, which corresponds to the beginning of the next phase given as $T_0$. 

Scaler value \text{$E$} is defined as 
\begin{equation} \label{Energy}
E=\int_{T_0}^{T_{end}}\pmb{\tau}_{in}^\top(t) \pmb{\omega}(t) \, \,dt
\end{equation}

Scaler value $T$ represents the duration of each phase, given as:
\begin{equation} \label{Tend}
T=T_{end}-T_0
\end{equation}
For the science phase, $T$ is given in table \ref{Objects}. $T$ is a variable defined by ML for the transient phase.

The constraints in SA are the constraints related to the dynamics of the system, the unity norm of the $\pmb{q_0}$, and the range of the variables. $N$ and $U$ are Gaussian (with $\mu$, $\sigma$ for mean and standard deviation) uniform distributions, respectively.

Multi-objectives Genetic Algorithm (MOGA) and SA find the optimal variables in the given unified (for both phases) non-convex constrained nonlinear optimization:

\begin{equation} \label{UnifiedFormulation}
\min_{\pmb{k}} \chi(\pmb{x}_0, w, \pmb{x}_f, \pmb{k})
\end{equation}
\noindent
Subject to the spacecraft rigid-body dynamics:
\begin{equation} \label{UnifiedDynamics}
\dot{\pmb{x}} = f(\pmb{x}(t),\pmb{\tau})
\end{equation}

\begin{equation*} \label{G*}
s.t:\hat{\pmb{\tau}}=G(\pmb{k},\pmb{x}(t))
\end{equation*}
\noindent
Quaternion normalization:
\begin{equation} \label{UnifiedQuatNorm}
\pmb{q}_0^T\pmb{q}_0 = 1
\end{equation}
\noindent
Control parameter bounds:
\begin{equation} \label{UnifiedParamBounds}
0.01 < k < 3 \quad (k \in \pmb{k}) 
\end{equation}
\noindent
Time constraint for the transient phase duration $(T)$:
\begin{equation} \label{UnifiedParamBounds}
7.2s < T < 72s
\end{equation}
\noindent
Final state bounds:
\begin{equation} \label{UnifiedFinalState}
\pmb{x}_f^l < \pmb{x}_f < \pmb{x}_f^u
\end{equation}
\noindent
Initial state bounds:
\begin{equation} \label{UnifiedInitialState}
\pmb{x}_0^l < \pmb{x}_0 < \pmb{x}_0^u
\end{equation}
\noindent
Initial condition distributions for the transient phase:
\begin{equation*} \label{Formulation412}
x_{0j}\sim U 
\quad (j=1,2,3,4)
\end{equation*}

\begin{equation*} \label{Formulation6}
x_{0j} \sim N(\mu_1,\sigma_1) \quad (j=5,6,7)
\end{equation*}
\noindent
\noindent
Final condition distributions:
\begin{equation*} \label{Formulation4}
x_{fj}\sim U 
\quad (j=1,2,3,4)
\end{equation*}

\begin{equation*} \label{Formulation61}
x_{fj} \sim N(\mu_2,\sigma_2) \quad (j=5,6,7)
\end{equation*}
Weight distribution for the transient phase:
\begin{equation} \label{UnifiedWeightDist}
w \sim \left[ N(\mu_3,\sigma_3), U(c_1^l, c_1^u), U(c_2^l, c_2^u) \right]
\end{equation}
Initial condition distributions for the science phase:
\begin{subequations} \label{UnifiedInitDist}
\begin{equation} \label{Constheta}
\pmb{q}_0=\pmb{\partial{q}}(\pmb{\theta}_0) \otimes \pmb{q}_f
\end{equation}

\begin{equation*} \label{Formulation63MGA}
\pmb{\theta}_{0} \sim N(\mu_4,\sigma_4) 
\end{equation*}

\begin{equation*} \label{Formulation6MGA}
\pmb{\omega}_{0} \sim N(\pmb{\mu}_j,\pmb{\sigma}_j)  \quad (j=5,6,7)
\end{equation*}

\begin{equation}
\pmb{x}_{0} = \begin{bmatrix} \pmb{q}_0\\\pmb{\omega}_{0} \end{bmatrix}
\end{equation}

\end{subequations}

\noindent



\noindent


The distribution of $\pmb{\omega}_0$ for the science phase is zero mean with the standard deviation chosen based on the table \ref{omegaFStat} and figure \ref{OmegaFinal}, as they show the $\pmb{\omega}_f$ distribution for the end of transient phase. The variance for the zero mean $\pmb{\theta}_0$ is chosen considering the $P_{99}(e)$ from the transient phase. $P_{99}(e)$ from the transient phase is chosen as the 3-sigma for the standard deviation of $\pmb{q}_0$ for the science phase.

The bounds of $\pmb{x_0}$ are

\begin{equation} \label{UpperLowerX}
\pmb{x_0}^l=\begin{bmatrix}
  -1\\-1\\-1\\0\\-2\\-2\\-2
\end{bmatrix}, 
\pmb{x_0}^u=\begin{bmatrix}
  1\\1\\1\\1\\2\\2\\2
\end{bmatrix}
\end{equation}

\begin{equation} \label{UpperLowerX1}
\pmb{x_f}^l=\begin{bmatrix}
  -1\\-1\\-1\\0\\0\\0\\0
\end{bmatrix}, 
\pmb{x_f}^u=\begin{bmatrix}
  1\\1\\1\\1\\0\\0\\0
\end{bmatrix}
\end{equation}

Lyapunov function control parameters $k_1$, and $k_2$ are positive scalar for global asymptotic stability.  The $U$ distribution bounds with the Gaussian in parameters are 
 

\begin{equation} \label{Uul2}
\pmb{c}^l=\begin{bmatrix}
0\\1
\end{bmatrix}, 
\pmb{c}^u=\begin{bmatrix}
  5\\1
\end{bmatrix},
\pmb{\mu}=\begin{bmatrix}
  0\\0\\5\\0\\0\\0\\0
\end{bmatrix}, 
\pmb{\sigma}=\begin{bmatrix}
  0.6\\0\\0.3\\0.0289\\0.052\\0.048\\0.058
\end{bmatrix}
\end{equation}
For a baseline mission with a 22.5-hour formation stabilization phase, $\omega_3$ reaches 0.59 rad/sec. Therefore, 0.6 is chosen as the 1-sigma of $\pmb{\omega}$ with zero mean distribution as given in equation \eqref{Uul2}. The initial quaternion distribution of the transient phase $\pmb{q}_0$ is a uniform distribution between -1 and 1 in equation \eqref{UpperLowerX} as the quaternion distribution varies between -1 and 1 for the formation phase. 

For the transient phase, SA solves the optimization equation (\ref{UnifiedFormulation}) and produces the data given in the table \ref{SAOAoutputStatisticsDNN}. MOGA solves the optimization equation (\ref{UnifiedFormulation}) for the science phase. 

\subsection{ML phase}

\subsubsection{Transient phase}

DNN is trained on 7893 data points produced in the data production phase. 
Both mean absolute percentage error (MAPE) and mean absolute error (MSE) are used for showing the estimation accuracy. k-fold cross validation with $ \text{k}=4$ is used to measure MAPE and MSE. As a result, 25\% of the data is used as test data. The inputs of the DNN belong to $\mathbb{R}^8$, and they are $\pmb{q}_0$, $\pmb{\omega}_0$, and $w$. 
The outputs of the DNN belong to $\mathbb{R}^5$ , and they are \text{$E$}, $e$, $T$ ,$k_1$, and $k_2$.

 Since all the outputs are positive, the ReLU \cite{pirayeshshirazinezhad2022designing,pirayeshshirazinezhad2022artificial} is chosen as the activation function for the output neurons. The hyperparameter of the DNN's structure are number of layers, number of neurons at each layer, and activation function. The parameters weight and bias of the DNN are optimized through the Adaptive Moment Estimation (Adam) \cite{kingma2014adam} and Nesterov-accelerated Adaptive Moment Estimation (Nadam) \cite{dozat2016incorporating} optimizer algorithms.

The optimization criterion consists of minimizing the MAPE through the Adam and Nadam algorithms.

The k-fold cross validation with $ \text{k}=4 $ is used to measure MAPE and MSE. As a result, 25\% of the data is used as the test data.  While training the DNN, 20\% of the training data is used as a validation data set for the early stopping. Number of batches, number of epochs, $l1$ regularization, kernel constraint are cross validated. For kernel constraint, $MaxNorm(x)$ function is used which limits the maximum norm of the weight vector for the layer with $x$. DNN's weights are bounded using $MaxNorm(x)$ function for kernel constraint.

The hyperparameter optimization problem obtains the minimum MAPE during the validation phase.
The hyperparameters of the DNN's structure and algorithm parameters are: number of layers, number of neurons at each layer, epochs, batches, weight initialization, $l1$ regularization, kernel constraint, activation function for the hidden layers, and patience. Adding other regularization methods including batch normalization and dropout doesn't reduce $\text{MAPE}$.

The coarse-to-fine optimization approach is to solve the hyperparameter optimization problem
first with GS. Next, the optimal solution
from GS is used to solve the hyperparameter optimization problem
with RS, which is finally validated on ALCF. The final optimal hyperparameter has 3 layers for each output. Other validated hyperparameter are given in \cite{pirayeshshirazinezhad2022designing}

\subsubsection{Science phase}
For this phase the controllers SMC and PD with MEKF are used to provide the leader and follower with arcminute accuracy. The star tracker and gyro sensor provide the attitude and angular velocity measurements for the MEKF. The trained DNN minimises the $e$ and $E$ in the science observation phase. 
DNN multi output is used for PD controller and SMC with 10 neurons in the first layer, 5 neurons in the second layer, and 3 neurons in the last layer. In the neural network in the training phase, the number of maximum epochs is set to 1000. Sigmoid is used as the activation function, and stochastic gradient descent (SGD) is used as the optimizer algorithm with MSE loss function.

\section{Experiments} \label{EXP}

\begin{itemize}
  \item The Monte Carlo simulation is used to analyse the performance and stability of the closed loop system.
    \item Monte Carlo simulation produces the data required for the ML.
  \item Monte Carlo simulation is used to design controller gains and $T$ using the mode of data set.

\end{itemize}


The ACS solver uses the fourth-order Runge-Kutta method (RK4) with a fixed time step to provide accurate and timely solutions for the ACS equations, ensuring that the quaternion unity norm constraint is maintained. The time step is chosen to stabilize 97\% of the data within a reasonable timeframe, with any remaining unstable solutions addressed by 300 iterations of Simulated Annealing (SA), resulting in an average data production time ($T_Q$) of 0.51 hours. This approach prioritizes accuracy and computational efficiency, removing the top 80\% of data with the highest $\chi$ values to enhance the accuracy of ML estimation.

\section{Data analysis and ML for transient phase}

\begin{figure}[!ht]
\centering
\includegraphics[width=0.74\textwidth]{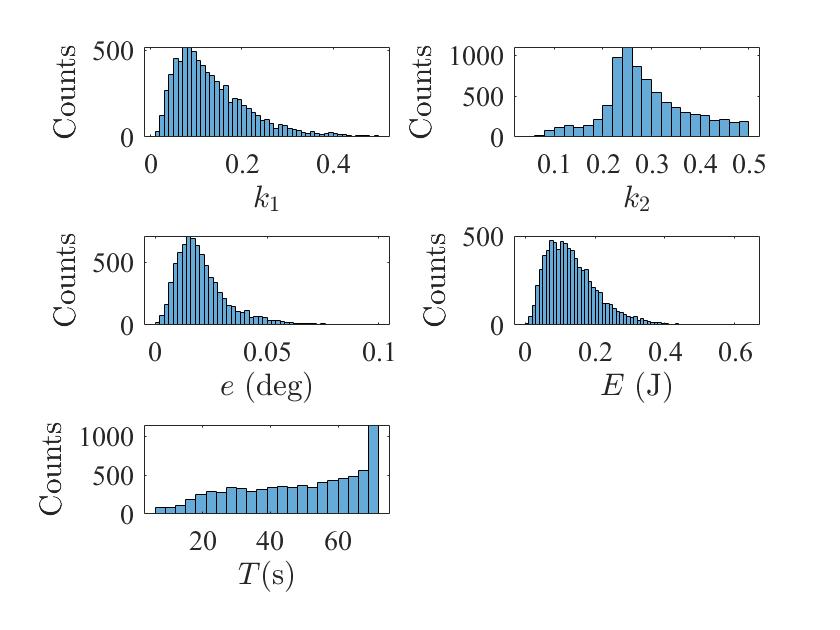}
\caption{Distributions of $k_1$, $k_2$, $e$, $E$, and $T$ from the data production phase. 
The parameters $k_1$, $k_2$, $e$, and $E$ all exhibit Gamma-like distributions. Figure adapted from \cite{pirayeshshirazinezhad2022designing}.}
\label{outputML}
\end{figure}

Fig. \ref{outputML} shows that the data are positive and bounded. The mode of the data set is between the maximum and minimum close to average except for $T$, which has its mode equal to its maximum. This data distribution shows that the $T$ is constrained and not minimized as its mode is close to its maximum. For $e$ and $E$ the mode is close to the average as they are minimized in the objective function.

\begin{table}[h!] 
\centering
\caption{Data statistics with 99 percentile values of DNN predicted outputs. }
\label{SAOAoutputStatisticsDNN}
\setlength{\tabcolsep}{3pt}
\begin{tabular}{|p{42pt}|p{53pt}|p{43pt}|p{57pt}|p{45pt}|p{50pt}|p{50pt}|p{55pt}|}
\hline
 $\pmb{y}$& Mean & Mode & Variance & $P_{1}$ & Max & $P_{99}$  & DNN ($P_{99}$)
\\
\hline
$k_1$ & 0.1342 & 0.085 & 0.0070 & 0.0244 & 0.4998 & 0.41 & 0.5627   \\\hline
$k_2$ & 0.2906 & 0.25 & 0.0077  & 0.0926 & 0.4998 & 0.49 & 0.7580 \\\hline
$e$ (deg) & 0.0233 & 0.015 & 6.2353e-04 & 0.0035 & 0.0990 & 0.067 & 0.0867 \\\hline
\text{$E$} (J)& 0.1349 & 0.075 & 0.0058 & 0.0229 & 0.6302 & 0.37 & 0.4695 \\\hline
$T$ (s) & 48 & 72 & 326.46 & 9 & 72.0 & 70.53 & 129.26 \\\hline
\end{tabular}
\end{table}

\begin{table}[h!] \centering
\caption{$\omega_f$ statistics.}
\label{omegaFStat}
\setlength{\tabcolsep}{3pt}
\begin{tabular}{|p{72pt}|p{53pt}|p{43pt}|p{57pt}|p{50pt}|p{45pt}|p{55pt}|}
\hline
 $\pmb{y}$& 
Mean & Variance & $P_{99}$  & $P_{1}$ 
\\
\hline






$\omega_{f1}$ (deg/s) & 1.03e-04   & 0.0027 & 0.1164 & -0.0972 \\\hline

$\omega_{f2}$ (deg/s) & -5.88e-04   & 0.0023 &  0.0592 & -0.1927 \\\hline

$\omega_{f3}$ (deg/s) & 9.08e-5  & 0.0034 &  0.1211 & -0.0938 \\\hline

\end{tabular}
\end{table}
 

\begin{figure}[!ht]
\centering
\includegraphics[width=0.74\textwidth]{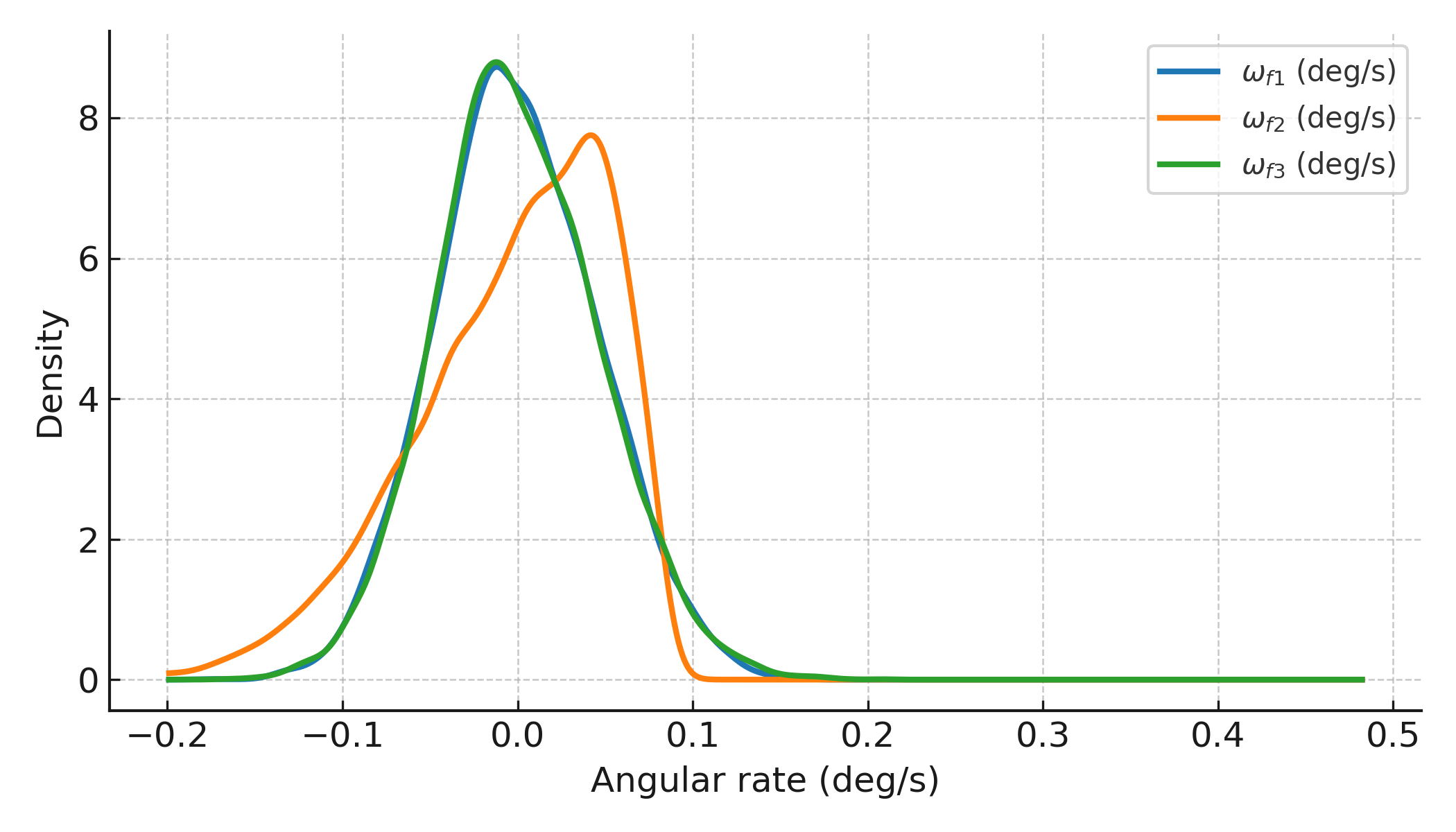}
\caption{ A kernel density estimate (KDE) for $\omega_f$ distribution. The distribution shows $\omega_f$ is centered around zero with a small variance.}
\label{OmegaFinal}
\end{figure} 

Table \ref{SAOAoutputStatisticsDNN} compares the statistics of the data with the prediction values from DNN. It highlights that the maxima of $e$ and $E$ are considerably higher than their $P_{99}$ values. 
This discrepancy comes from solver sensitivity in the ACS, noise in the dynamics, the stochastic nature of simulated annealing, and the heavy-tailed character of the resulting distributions.
As $e$ converges to zero, the terminal angular velocity $\omega_f$ also converges toward zero, consistent with the histogram in Fig.~\ref{OmegaFinal}. 
The $P_{99}$ and $P_{1}$ bounds for all $\omega_f$ components lie within $\pm 1~\mathrm{deg/s}$, confirming tight dispersion around zero.

The percentile statistics demonstrate that mission requirements are satisfied by the optimization data: 
$P_{99}(e)=0.067^\circ < 0.09^\circ$ and $P_{99}(E)=0.37$ J per orbit. 
This energy budget remains comfortably low for long-duration operations, well within current spacecraft technology capabilities. 
In this sense, the simulated annealing optimization plays a role similar to model predictive control (MPC): actively tuning gains to satisfy strict constraints. 
The drawback is computational cost—on average, about 30 minutes are required to complete a single optimization run.

\subsection{Transient phase ACS stability analysis}

In the ML approach, using ReLU as the activation function for the output layer forces the controller gains $k_1$ and $k_2$ to be positive scalars\eqref{UnifiedParamBounds}. Since ML defines the positive Lyapunov controller gains at the beginning of the transient phase, the controller is globally asymptotic stable without considering the noise, disturbances, and model uncertainties. However, the closed loop system is shown to be stable using Monte-Carlo simulation since $P_{99}$(\text{$e$}) and $P_{99}$(\text{$E$}) are bounded.
The closed loop system is stable and $P_{99}$(\text{$e$}) $<$0.2 deg and $P_{99}$(\text{$E$})$<$2 J. Having the kernel constraint enforces the weights to be bounded so the DNN outputs are bounded when the inputs are bounded. Besides, the ML gives $P_{99}$(\text{$\hat{e}$}) $<$0.09 deg and $P_{99}$(\text{$\hat{E}$})$<$0.5 J using Monte Carlo simulation for all the data, which shows the stability of closed loop system using DNN.

\section{Data analysis and ML for science phase} \label{EXPS}
\begin{figure}[!ht]
\centering
\includegraphics[width=0.74\textwidth]{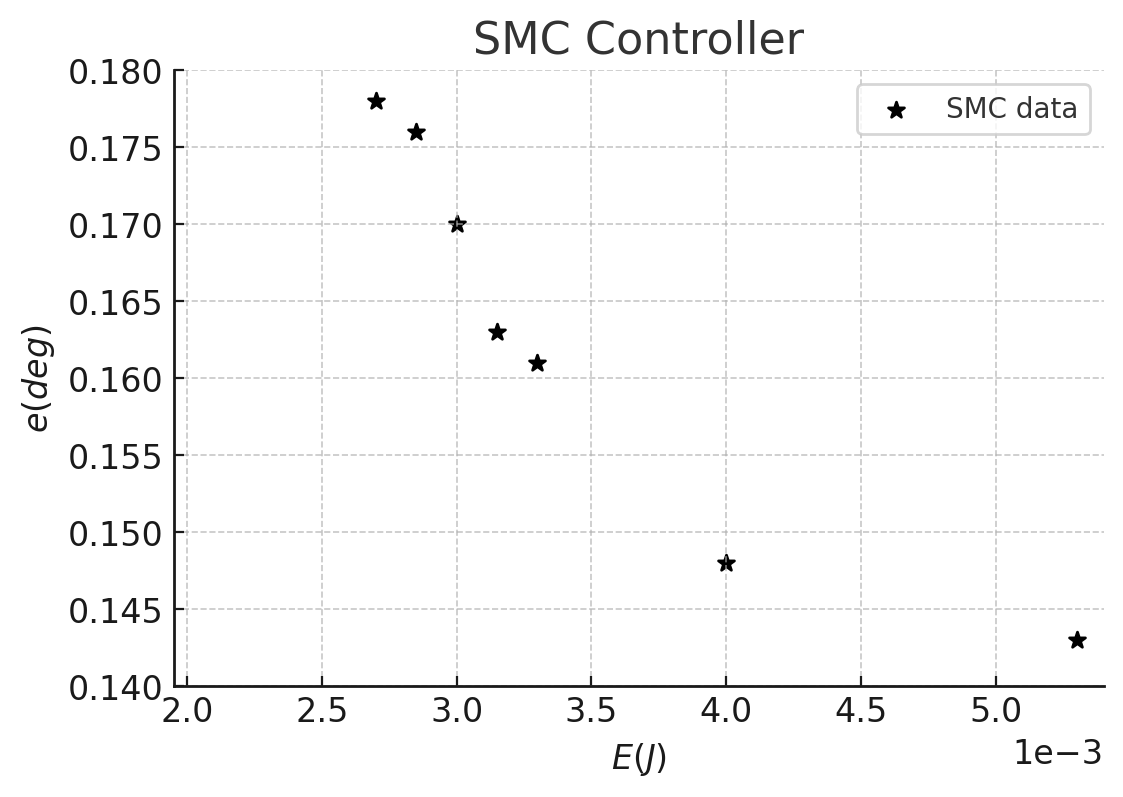}
\caption{The Pareto front for the sliding mode controller in the science observation phase.}
\label{SMCGAP}
\end{figure}

The data is produced using multi objective genetic algorithm (MOGA) for SMC and PD controllers. One sample of MOGA Pareto front for SMC is shown in Fig \ref{SMCGAP}.

\subsection{ML results}

When using DNN, the linear controller PD and SMC provide the MSE as 0.0434 and 0.0124, respectively. These ML are used to predict the controller parameters $\pmb{k}$. SMC provides the VTXO with lower MSE compared to PD controller. 

\subsection{Simulation results}

Two experiments are done to show the performance of the given methodology. For two given $w$, the estimated optimal controller parameter by ML,  $\hat{\pmb{k}}$, is given in Table \ref{INOU}.
\begin{table}[ht!]
\caption{$\hat{\pmb{k}}$ values}
\setlength{\tabcolsep}{3pt}
\centering
\begin{tabular}{|p{90pt}|p{190pt}|}
\hline
$w$ & $\hat{\pmb{k}}$
\\
\hline
$w_{PD}=6.28$ & $\hat{\pmb{k}}_{PD}=[0.1208 \quad 0.3786]^T$
\\
\hline
$w_{SMC}=26.14$ &  $\hat{\pmb{k}}_{SMC}=[2.9947 \quad 0.0193 \quad 0.2601]^T$ \\\hline

\end{tabular}
\label{INOU}
\end{table}

Using $\hat{\pmb{k}}$ from the ML, the $e$ and $E$ are obtained using the GNC simulation given in Table \ref{IRNOU}.  

\begin{table}[ht!]
\caption{$e$ and $E$ obtained from $\hat{\pmb{k}}$.}
\setlength{\tabcolsep}{3pt}
\centering
\begin{tabular}{|p{60pt}|p{60pt}|p{60pt}|}
\hline
controller & $e$ (deg) & $E$ (J)
\\
\hline
PD & 0.2219 & 0.0353
\\
\hline
SMC &  0.1738 & 0.0066 \\\hline

\end{tabular}
\label{IRNOU}
\end{table}
Table. \ref{IRNOU} shows that $e$ is a few arcminute which satisfies the mission requirement for each individual spacecraft. In this example, PD controller error $e$ here is more than 0.18 deg, which is not acceptable. SMC is used as it provides a more robust controller toward disturbances and uncertainties for the nonlinear spacecraft dynamics, with a low prediction error as 0.0124.

\section{Relative position formation} \label{RelP}
The relative position formation is shown for the science observation phase. For simplicity, it is assumed the leader is in its natural orbit and the follower controls the relative position. Disturbances are included in the dynamics given in section \ref{Dist}. SMC and PD controller are given in section \ref{Control}. The sensors used for measurements are the radio ranging for relative distance measurements $\tilde{r}^z_{rel}$, IMU the for relative velocity measurements $\tilde{\pmb{\text{v}}}_{rel}$, and the interferometry sensor for transverse alignment measurements $\tilde{\pmb{r}}^{xy}_{rel}$. As a result, each sensor measures the states of the dynamics individually. The sensor measurements are included in the simulations. The parameters for the PD controller is chose as $P=1$ and $D=1$. For the SMC, the sat function is chosen as a sign function for simplicity, and the controller parameters are $k=1$ and $\pmb{Z}=1$ for each axis. The desired relative velocity and the desired relative position are

\begin{equation} \label{Des}
\pmb{\text{v}}^f_{rel}=0_{3\times1}
\end{equation}

\begin{equation} \label{DesRe}
\pmb{r}^f_{rel}=\begin{bmatrix}
  0 \\ 0 \\ 1 
\end{bmatrix} (\text{km})
\end{equation}

The initial conditions are


\begin{equation} \label{DesIn}
\pmb{\text{v}}^0_{rel}=\begin{bmatrix}
  1 \\ 1 \\ 1 
\end{bmatrix}(\text{m/s})
\end{equation}

\begin{equation} \label{DesReIn}
\pmb{r}^0_{rel}=\begin{bmatrix}
  1 \\ 2 \\ 10 
\end{bmatrix}(\text{km})
\end{equation}

\begin{figure}[!ht]
\centering
\includegraphics[width=0.74\textwidth]{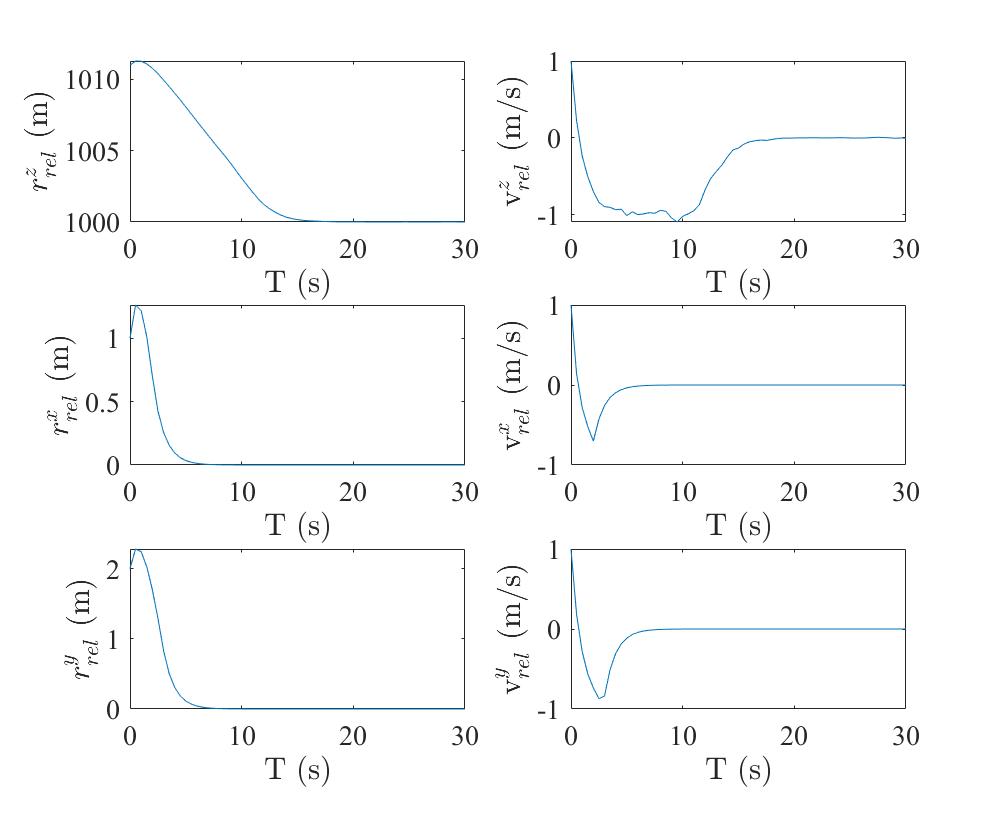}
\caption{The relative position formation using SMC.}

\label{SMCRel}
\end{figure} 

\begin{figure}[!ht]
\centering
\includegraphics[width=0.74\textwidth]{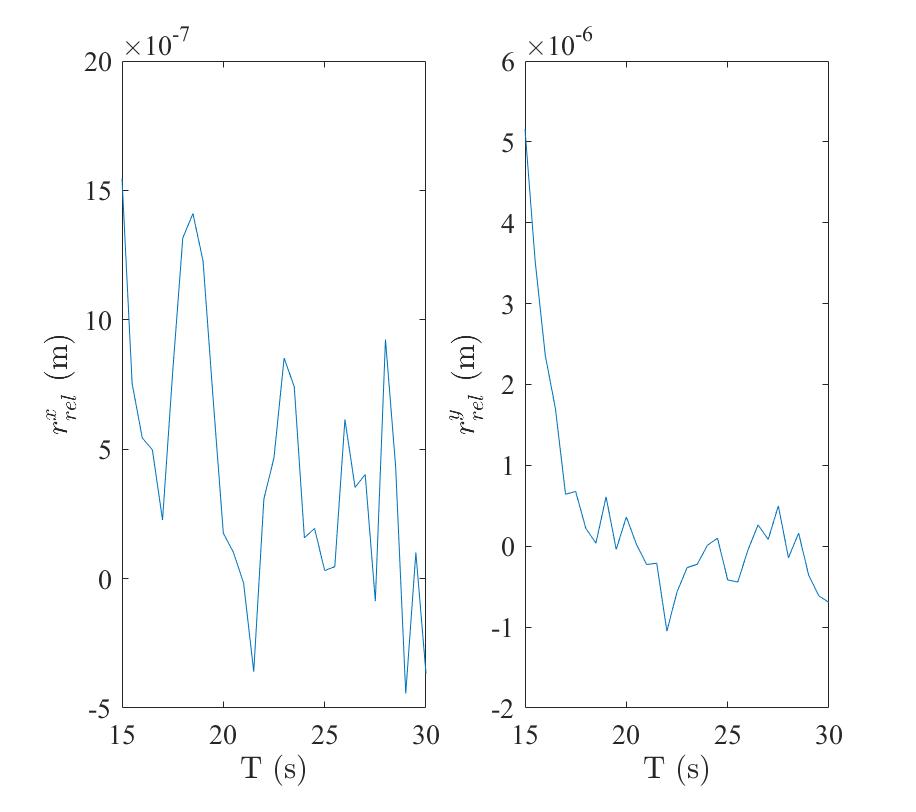}
\caption{Transverse alignment using SMC when the system is settled for the last 15 s.}

\label{SMCRelLast}
\end{figure}

The transverse alignment using SMC for the last 15 s is shown Fig. \ref{SMCRelLast}

Fig. \ref{SMCRel} shows that the system converges to the desired trajectory. Fig. \ref{SMCRelLast} shows that relative position alignment stays in the order of $10^-6$, which satisfies the sub millimeter accuracy. The transverse alignment fluctuates because of the noise in the sensors and disturbances in the system.



\begin{figure}[!ht]
\centering
\includegraphics[width=0.84\textwidth]{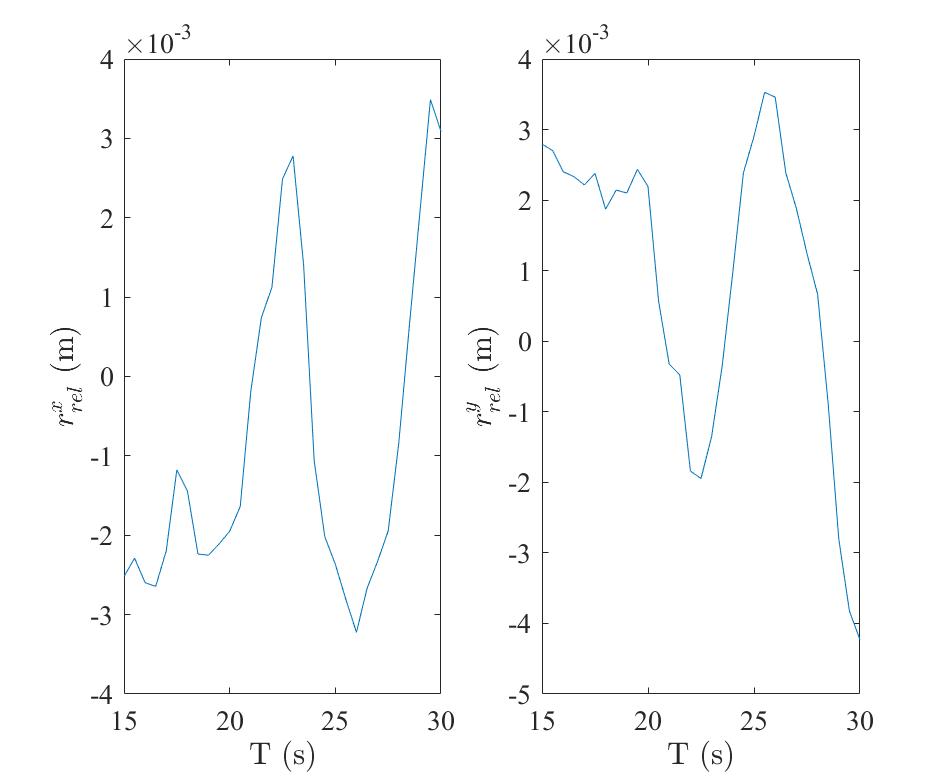}
\caption{Transverse alignment using PD when the system is settled for the last 15 s.}

\label{PDRelLast}
\end{figure}

The transverse alignment using PD for the last 15 s is shown Fig. \ref{PDRelLast}.

The PD controller makes the system converge to the desired trajectory. Fig. \ref{PDRelLast} shows that relative position alignment stays in the order of $10^-3$, which doesn't satisfy the sub millimeter accuracy. The transverse alignment fluctuates because of the noise in the sensors and disturbances in the system. It shows that the designed PD controller can't satisfy the science phase accuracy requirement. However, the PD can be modified to PID or a better tuned PID controller for a better tracking characteristic. The added integral to the controller corrects for steady-state error by accumulating the error over time.

\section{Supervisory adaptive control system for ACS}

The timed automata method is used to model the supervisory adaptive control system \cite{hendriks2006model,ghorashi2021distributed}. The stability of hybrid control systems are discussed in \cite{paxman2004switching}. The system includes different phases and controllers. The timed automata method models the changes in the phases with the needed conditions for it. The commissioning, ACS phases in collaboration with the relative position control, and de-commissioning are expressed below.

The commissioning takes for 60 days. After the commissioning, the first space object begins to be observed in the science phase after the transient phase and when the relative distance is achieved with 1 meter accuracy. In the science observation phase, the formation holds for the duration given in Table \ref{Objects1} for each space object, after the commissioning and before the de-commissioning.  In Table \ref{Objects1}, the space objects, the commissioning, and the de-commissioning are considered as the operational mode. The Id shows the operational mode index.

\begin{table}[ht!]
\caption{Finite automata rule $\pmb{u}_3$.}
\setlength{\tabcolsep}{3pt}
\centering
\begin{tabular}{|p{40pt}|p{170pt}|p{55pt}|}
\hline
Id & Operational mode state $s_4$ & 
Rule $\pmb{u_3}$
\\
\hline
0 & Commisioning & 60 days \\\hline
1 & Sco X-1 & 0.2 hr \\\hline

2 & GX 5-1 & 1.5 hr \\\hline

3 & GRS 1915+105 & 4.2 hr  \\\hline

4 & Cyg X-3 & 4.9 hr\\\hline

5 & Crab Pulsar & 5.4 hr \\\hline

6 & Cen X-3 & 19 hr\\\hline

7 & $\gamma$Cas & 146 hr \\\hline

8 & Eta Carinae & 452 hr\\\hline

9 & De-commissioning & 5 days\\\hline

\end{tabular}
\label{Objects1}
\end{table}

After the science observation phase, the formation stabilization phase completes the 32.5 hr duration of the orbit if still observing the same object for the duration given in Table \ref{Objects1}.

This cycle of phases repeats until the observation period for the objects given in Table \ref{Objects1} is met, and then the orbits are switched to the new orbits for the new space objects in Table \ref{Objects1}, and the cycle repeats for the new space object.  After the mission is completes, the de-commissioning deorbits the spacecraft.

The timed automata model of the designed supervisory control system is a 5-tuple as
$\pmb{TA}=\{\pmb{s},\pmb{t},\pmb{i},\pmb{k},\pmb{Init}\}$. Each tuple is explained below.

\begin{itemize}

  \item $\pmb{s}=\{s_0, s_1, s_2, s_3, s_4, s_5\}$ is the set of discrete states of the finite automata model of the system. Each discrete state of the system $s_i$ is associated with a specific phase or an operational model of the system.
  
\item $\pmb{t}=\{t_1, t_2, t_3\}$ is the set of local timers. The local timer shows the amount of time that the system has spent in each state $\pmb{s}$.

\item $\pmb{i}=\{u_1, u_2, u_3, u_4\}$ is the set of the inputs to supervisory control system. 

\item $\pmb{k}=\{ \pmb{k_1}, \pmb{k_2}, \pmb{k_3}\}$ is the set of all transitions. A transition $k_i$ is a four tuple given by
\begin{equation} \label{Transition}
\pmb{k}_{ij}=\{s_i,g(i,j),reset(i,j),s_j\}
\end{equation}

$s_i$ and $s_j$ are the source and destination states, and $g(i,j)$ is the guard condition for the transition. Gaurd conditions are Boolean expressions that are evaluated based on the inputs $\pmb{i}$ and local timer $\pmb{t}$. Table \ref{GC} shows the guard conditions and their corresponding Boolean expressions in the designed supervisory control system. $reset(i,j):R\rightarrow R$ is the reset condition for the transition.

\item $\pmb{Init}=\{c_1, c_2, c_3\}$ is the set of initial conditions for the system where $\pmb{Init}\subset[\pmb{s},\pmb{t}]$

\end{itemize}

In the design of the architecture of the timed automata model, the set of discrete states of the finite automata $\pmb{s}$ is given as below
\begin{itemize}

\item $s_0$: Commissioning.
  \item $s_1$: Formation stabilization phase.
    \item $s_2$: Transient phase.
      \item $s_3$: Science observation phase.
        \item $s_4$: Next operational mode state. At this state, we go to the next operational mode, and the Id in Tbale \ref{Objects1} is increased.
                \item $s_5$: De-commissioning.
\item $s_6$: End of operation.
        \end{itemize}

In the design of the architecture of the timed automata model, the set of local timers $\pmb{t}$ is given as below
\begin{itemize}

  \item $t_0$: Commissioning duration.
  \item $t_1$: Formation stabilization phase time.
    \item $t_2$: Transient phase time.
      \item $t_3$: science phase time.
        \item $t_4$: Next operational mode time.
                \item $t_5$: De-commissioning duration.
        \end{itemize}
        
In the design of the architecture of the timed automata model, the set of the inputs to supervisory control system $\pmb{i}$ is given as below:

\begin{itemize}
  \item $u_1$: Formation stabilization phase rule. It is obtained by subtracting the 10 hr duration of the science phase and the maximum 3 min duration of the transient phase from the total orbit period. As a result, $u_1=32.5 \text{hr}-10 \text{hr}- 3\text{min} $.
    \item $u_2$: Transient phase rule, which is $u_2=T$ given by the ML.
      \item $u_3$: Duration rule from the Table \ref{Objects1} for the operational mode $s_4$. $\pmb{u_3}=[u_3(Id)]$ where Id shows the updated operational mode in state $s_4$. Initially, $u_3(Id=0)$ corresponds to the duration of commissioning which is 60 days. 
        \item $u_4$: The duration of reaching the science observation in the relative position control.
        \end{itemize}
        
In the design of the architecture of the timed automata model, the set of all transitions $\pmb{k}$ is given as below

\begin{itemize}

  \item $k_0$: Commissioning transition law. 
    \item $k_1$: Formation stabilization phase transition law.
      \item $k_2$: Transient phase transition law.
        \item $k_3$: science phase rule, De-commissioning transition law. 
          \item $k_4$: ML rule, next operational mode transition law.
                    \item $k_5$: End of operation transition law.

        \end{itemize}

In the design of the architecture of the timed automata model, the set of initial conditions $\pmb{Init}$ is given as below

\begin{itemize}

  \item $c_0$: Commissioning initial conditions. 
    \item $c_1$: Formation stabilization phase initial conditions.
      \item $c_2$: Transient phase initial conditions.
        \item $c_3$: science phase initial conditions.
          \item $c_4$: Next operational mode initial conditions.
                    \item $c_5$: De-commissioning initial conditions.
                     \item $c_6$: End of operation initial conditions.
        \end{itemize}
$\pmb{Init}:[\pmb{s}=\pmb{s}(0),t_i=0 (i=0,1,...,6)]$, which corresponds to all the local timers set to zero at the beginning, and $\pmb{s}$ are at their initial condition.
The initial condition of $s_4$ corresponds to the commissioning in the operational mode which is $Id=0$.

\begin{table}[ht!]
\caption{Reset function $reset(i,j)$.}
\setlength{\tabcolsep}{3pt}
\centering
\begin{tabular}{|p{60pt}|p{60pt}|}
\hline
$reset(i,j)$ & Operation
\\
\hline
$reset(0,4)$ &  $t_0=0$ \\\hline
$reset(1,2)$ &  $t_1=0$ \\\hline
$reset(2,3)$ &  $t_2=0$ \\\hline
$reset(3,4)$ &  $t_1=0$\\\hline
$reset(3,5)$ &  $t_3=0$\\\hline
$reset(4,2)$ &  $t_1=0$\\\hline
$reset(5,6)$ &  $t_5=0$\\\hline
\end{tabular}
\label{ResetC}
\end{table}

\begin{table}[ht!]
\caption{Boolean guard conditions $g(i,j)$.}
\setlength{\tabcolsep}{3pt}
\centering
\begin{tabular}{|p{120pt}|p{110pt}|}
\hline
Guard condition $g(i,j)$& 
Boolean expression
\\
\hline
$g(0,4)$ &  $u_3<t_0$ \\\hline
$g(1,2)$ &  $u_1<t_1$ \\\hline
$g(2,3)$ &  $u_2<t_2$ \\\hline
$g(3,4)$ &  $u_3 \geq t_3$\\\hline
$g(3,1)$ &  $u_3<t_3$\\\hline
$g(3,5)$ &  $452 \text{hr} \leq t_3$\\\hline
$g(4,2)$ &  $u_4<t_4$\\\hline
$g(5,6)$ &  $u_3<t_5$\\\hline

\end{tabular}
\label{GC}
\end{table}

\begin{figure}[!ht]
\centering
\includegraphics[width=0.8\textwidth]{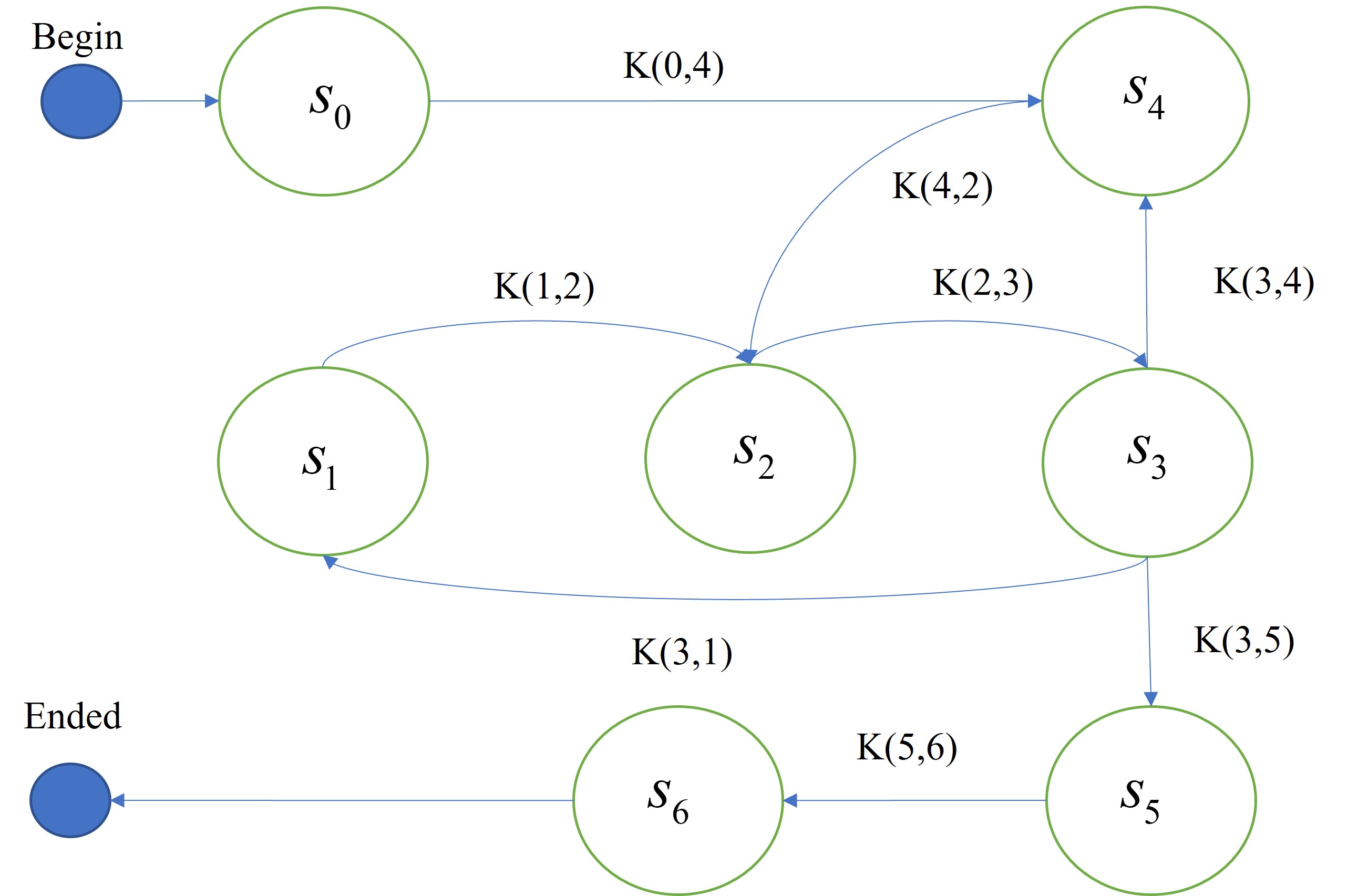}
\caption{Supervisory adaptive control system. $\pmb{s}$ represents the set of discrete states of finite automata, and $\pmb{k}$ represents the set of all transitions in finite automata}
\label{ST}
\end{figure}

Fig. \ref{ST} illustrates the supervisory adaptive control system.

In the supervisory control system, ML is used for updating $T$ and controller parameters in GNC to make the supervisory control system intelligent. The intelligent supervisory control system increases the lifetime of the mission and satisfies the requirements of the mission. SMC provides global asymptotic stability with positive controller parameters for any initial condition and any desired trajectory. In the intelligent supervisory control system, the initial conditions, the desired trajectory, and controller parameters are updating while the system dynamics remains the same. As a result, When SMC is used for the GNC, the supervisory control system provides asymptotic stability. The global asymptotic stability of SMC are proved in sections \ref{SMCACS},\ref{SMCR} and \ref{SMCL}.




\section{Verification and Validation (V\&V) of VTXO mission}

\subsection{Domain and objective}

The main objective of the V\&V section is to confirm the accuracy, reliability, and efficiency of the control systems implemented for the VTXO space mission. The V\&V process aims to ensure that the systems perform consistently under different environmental conditions and adhere to the defined mission objectives.

\subsection{Methods and Approach}

This section establishes that the VTXO guidance, navigation, and control (GNC) architecture - comprising a timed automatic supervisory policy, a Lyapunov-based transient controller with ML-selected gains, a sliding–mode controller (SMC) for science pointing and relative position regulation, and MEKF-based navigation; satisfies the mission's precision, energy, timing, and robustness requirements under realistic disturbances and sensing conditions. The verification approach combines formal stability arguments with large-scale closed-loop Monte Carlo (MC) simulation campaigns and distributional performance summaries derived from optimization-backed data generation.

We adopt a simulation-first methodology because the hybrid nature of the system couples continuous spacecraft dynamics with discrete supervisory transitions. All closed-loop simulations are performed with a fixed-step fourth-order Runge–Kutta (RK4) integrator, which preserves the quaternion norm and provides stable error control across phases. Sensor noise, inertia modeling errors, reaction-wheel nonlinearities, and external torques are injected according to the parameterizations given earlier (Tables~\ref{RWParam}, \ref{Uul2}, \ref{Keplerian elements}), so that navigation, control, and supervisory decisions are exercised under the same uncertainties expected on-orbit. Phase progression is governed by a timed–automata policy (Fig.~\ref{ST}) with explicitly defined guards and timer resets (Tables~\ref{GC}–\ref{ResetC}). This ensures non-Zeno behavior and, more importantly for correctness, enforces that the reachable states at each phase boundary lie inside the admissible set for the next phase. In particular, the transition from the transient phase to the science phase triggers only when the ML-predicted duration $T$ is attained and the terminal accuracy is within tolerance; the supervisor therefore mediates a safe hand-off from aggressive slews to persistent precision tracking.

The controllers are verified at two levels. At the analytical level, the transient controller’s global asymptotic stability follows from a standard Lyapunov argument when the feedback gains $(k_1,k_2)$ are positive. Within our framework these gains are produced by a DNN with a ReLU output layer and kernel max-norm constraints, which guarantees positivity and boundedness of the outputs for bounded inputs; thus the Lyapunov conditions are preserved at run time. For the science observation and relative position problems, the SMC laws adopt a saturation layer in place of the discontinuous sign function, eliminating high-frequency chattering while preserving the invariance of the sliding surface in the presence of bounded disturbances and model uncertainties. The SMC stability proof is given earlier in Sections~\ref{SMCL} and \ref{SMCR}, and those results transfer directly to the relative position formulation where the counterpart spacecraft’s thrust is treated as a disturbance.

At the empirical level, we validate performance with MC campaigns whose initial conditions, disturbances, and sensor characteristics follow the distributions used to generate the optimization dataset. The dataset itself is obtained with SA/MOGA solvers that minimize a transient-phase scalar loss and a science-phase vector loss $\chi=[E,\;e]^\top$, and its statistics are summarized in Table~\ref{SAOAoutputStatisticsDNN}. These summaries provide distributional guarantees that are more informative than single-scenario traces. In particular, the $99^{\text{th}}$ percentile values $P_{99}(e)=0.067^\circ$ and $P_{99}(E)=0.37$\,J demonstrate that extreme events are rare and that mission accuracy and energy budgets are met with high probability, while maxima are naturally higher due to solver stochasticity and the heavy-tailed character of the induced distributions. The terminal angular velocity $\omega_f$ distribution (Fig.~\ref{OmegaFinal}) is centered near zero with narrow $P_{1}$ and $P_{99}$ bounds within $\pm1$\,deg/s, aligning with the requirement that steady tracking proceeds without persistent rate bias. For the science observation phase, SMC maintains arc-minute attitude accuracy and sub-millimeter transverse alignment in the presence of injected noise and disturbances, as seen in the relative-position experiments (Figs.~\ref{SMCRel} and \ref{SMCRelLast}); the PD baseline under identical conditions tracks at the arc-minute level but does not achieve sub-millimeter alignment (Figs \ref{PDRelLast}), which corroborates the choice of SMC for flight.

A critical question for an AI-assisted controller is whether the ML layer jeopardizes stability or erodes margins. Here the learning component is deliberately confined to selecting positive controller gains and the transient duration $T$ from offline-optimized data reflecting the operational envelope. The DNN is trained on the SA/MOGA dataset with explicit regularization and max-norm constraints, and its outputs are further validated in MC: across folds and held-out conditions, the predicted pairs $(\hat{k}_1,\hat{k}_2)$ and durations $\hat{T}$ yield closed-loop responses that respect the same percentile criteria as the generating optimization runs. In other words, the ML layer inherits the feasibility and margins of the optimization dataset rather than attempting to learn a control policy ex nihilo. This design choice, together with the positivity guarantees furnished by the ReLU output layer, preserves the Lyapunov and SMC conditions by construction.

The supervisor itself is validated along two axes: liveness and safety. Liveness is guaranteed by guards that require strictly positive dwell times for each location, ruling out Zeno switching. Safety is demonstrated by showing that the guard conditions are only enabled when the next phase’s admissible set is reachable under the current controller. In practice this is evidenced by the distributional bounds at phase end: during the transient phase, the last-$D$-second error norm meets the $e$ tolerance at the ML-predicted $\hat{T}$, enabling the $g(2,3)$ guard; at science-phase exits, the $u_3$ timing rule and energy consumption remain within the planned envelope, enabling either a return to stabilization or a transition to the next operational mode. The resulting hybrid execution sequences align with the mission timeline in Table~\ref{Objects1} and complete without deadlocks.

The verification evidence therefore triangulates across formal arguments, distributional guarantees, and time-domain closed-loop behavior. Formal arguments ensure that, under the enforced positivity and boundedness constraints on gains and SMC parameters, the continuous dynamics are globally asymptotically stable in the transient phase and asymptotically converge to the sliding manifold in the science and relative position phases. Distributional guarantees ($P_{99}$ bounds for $e$, $E$, and $\omega_f$) demonstrate that robustness extends beyond nominal conditions to the broad support of the operational envelope. Time-domain evidence shows that under the same noise and disturbance models used for data generation, the system tracks scientific targets within the required arc-minute pointing and sub-millimeter alignment while consuming energy consistent with the optimization-derived budgets.

Two limitations merit explicit acknowledgement. First, all probabilistic statements are conditional on the disturbance, actuator, and sensor models employed. Changes in on-orbit environment or component aging that shift these distributions would call for retraining the ML map and regenerating the SA/MOGA dataset to re-establish the same percentile margins. Second, the heavy-tailed nature of the optimization outputs implies that rare outliers above $P_{99}$ can occur; operational practice mitigates these events by allocating margin in propellant and observation windows and by allowing the supervisor to prolong the transient phase when necessary.

Within these assumptions, the V\&V results support the claim that the VTXO GNC system, as architected here, meets mission requirements with high confidence. The hybrid execution orchestrated by the timed automaton, the stability and robustness of SMC for precision phases, and the constrained, explainable role of ML in parameter selection collectively yield a control stack that is both verifiably correct and practically deployable. Reproducibility is facilitated by reporting aggregated dataset statistics (Table~\ref{SAOAoutputStatisticsDNN}) and by referencing the distributional behavior of outputs and terminal rates (Figs.~\ref{outputML} and \ref{OmegaFinal}); these artifacts enable independent replication of the MC campaigns and cross-checking of the reported percentiles.

\section{Conclusion}

The VTXO mission's demanding requirements for high-precision formation flying are addressed through a hierarchical control architecture that partitions the problem into distinct operational phases. This work demonstrates that combining Simulated Annealing (SA) optimization with a Deep Neural Network (DNN) surrogate model provides a robust method for finding optimal control parameters that minimize energy consumption ($E$) and pointing error ($e$) within a constrained maneuver time ($T$). Given the significant computational cost of optimization (averaging 0.51 hours per run), the DNN serves as an efficient, real-time surrogate, learning the optimal behavior from the offline-generated data.

A pivotal contribution of this research is the development of an Explainable AI (XAI) framework. By training the DNN to predict not only the control gains but also the resultant $E$ and $e$, the system transcends the typical ``black-box'' paradigm. This predictive transparency allows the system's decisions to be interpretable and verifiable, a critical feature for mission-critical applications. This explainability is further reinforced by the timed automata-based supervisory control, which provides a clear, rule-based logic for all phase transitions.

The efficacy of this approach is validated through extensive Monte-Carlo simulations, which were used for both data generation and sensitivity analysis. SA optimization successfully balanced the competing mission objectives, and the hyperparameter-tuned DNN learned this behavior with high fidelity. The inclusion of noise and disturbances in the training data acts as a form of regularization, enhancing the robustness of the learned model.

The comprehensive Verification and Validation (V\&V) process confirms that the proposed control system is not only efficient and robust but also trustworthy. By rigorously testing the system across its full operational envelope, we have validated its ability to meet stringent mission objectives. This V\&V effort substantiates the reliability of the AI's predictions and the logical soundness of the supervisory framework, confirming its readiness for complex, autonomous space environments.
\appendix \label{Appendix}

\section{Asymptotic Stability}

A point is considered to be an equilibrium point $\pmb{x}_e$ for the system if $\dot{\pmb{x}}(t)=0$ for all $t$. 
$\pmb{x}_e$ for the system is global asymptotic stable if the positive scalar function $V(\pmb{x})$ satisfies the following conditions
\begin{itemize}

  \item $V(\pmb{x}_e)=0$
  \item $V(\pmb{x})>0$ for $\pmb{x}\neq \pmb{x}_e$ 
  \item $\dot V(\pmb{x})\leq0$
\end{itemize}
When the given conditions are satisfied, $V(\pmb{x})$ is a Lyapunov function. If $\dot V(\pmb{x})<0$ for $\pmb{x}\neq\pmb{x}_e$, $\pmb{x}_e$ is asymptotically stable. If $\dot V(\pmb{x})\leq0$, $V(\pmb{x})$ is a Lyapunov function and the system is stable. LaSalle’s theorem can prove the asymptotic stability.

\section{Sliding mode control asymptotic stability proof} \label{SMCL}

The SMC stability can be proven/shown by the Lyapunov stability theorem. In SMC, the state $x$ reach the desired state $x_f$. The difference between the states and the desired states is

\begin{equation} \label{Dx}
\Delta(x)=x-x_f
\end{equation}

Since the dynamics of spacecraft is second order, the Lyapunov asymptotic stability proof is shown for a second order system. A second order system is given by 

\begin{equation} \label{Second order}
\ddot{x}=f(x(t),\dot{x}(t))+u(t)
\end{equation}

A sliding surface is considered as

\begin{equation} \label{SlidingS}
s=\Delta\dot{x}+\lambda\Delta(x)
\end{equation}

Where $\lambda$ is a scalar. The following Lyapunov function proves the asymptotic stability 

\begin{equation} \label{vs}
V(\Delta(x))=\frac{1}{2}s^2
\end{equation}

The time derivative of the Lyapunov function is given as
\begin{equation} \label{vsd}
\dot V(x)=s\dot{s}
\end{equation}

The SMC control law is obtained by preventing the motion off of the sliding surface by setting $\dot{s}=0$. To obtain the control law, the known model is defined as $\bar{f}(x,\dot{x})$. Using $\dot{s}=0$ into \eqref{Second order} and taking the time derivative of s in \eqref{SlidingS}, $\dot{s}$ is obtained as

\begin{equation} \label{vsdU}
\dot s=\bar{f}(x,\dot{x}) +u -\ddot{x}+\lambda\Delta(\dot{x})
\end{equation}

The nominal input $\bar{u}$ that derives the $\dot s$ to zero is obtained as

\begin{equation} \label{Us}
\bar{u}=-\bar{f}(x,\dot{x}) +\ddot{x} - \lambda\Delta(\dot{x})
\end{equation}

Since there are model uncertainties and disturbances in the system, the discontinuous term $-ksign(s)$ is added to the $\bar{u}$ given by

\begin{equation} \label{Ureal}
u=\bar{u}-ksign(s)
\end{equation}

Assuming that model uncertainties and disturbances in the system are bounded by the known function $F(x,\dot{x})$ with a maximum value $F_{\text{max}}$, and letting $k=F_{\text{max}}+\varrho$ for the positive scalar $\varrho$, $\dot V(x)$ is obtained as
\begin{equation} \label{vsdUdot}
\dot V(x)\leq-\varrho|s|
\end{equation}

When $k$ is chosen to be large enough, the given sliding mode control law is stable in the presence of uncertainties and disturbances in the system. Larger $k$ introduces more chattering since $ksign(s)$ induces chattering the system. To reduce chattering, $sign(s)$ can be replaced by a saturation function with a varying boundary layer thickness, and the control law is given as

\begin{equation} \label{UrealChatter}
u=u-ksat(s,\varepsilon )
\end{equation}

$\varepsilon$ is the boundary layer thickness.

The proof of sliding mode control (SMC) stability are more discussed in \cite{markley2014fundamentals}.

\section{Sensor model VTXO}\label{}
\subsection{Gyro model}

In $\tilde{\pmb{\omega}}$, Gaussian white noise $\pmb{w}_{\omega}$ (angular random walk), bias $\pmb{b}_{\omega}$, scale factor $\pmb{f}_{\omega}$, and misalignment $\pmb{\epsilon}_{\omega}$ are included as the following

\begin{equation} \label{omegaMeasure}
\tilde{\pmb{\omega}}=\partial \pmb{R}(\pmb{\epsilon}_\omega)[\{\pmb{I}_{3\times3}+\text{diag}(\pmb{f}_\omega)\}\pmb{\omega}+\pmb{b}_\omega+\pmb{w}_\omega]
\end{equation}

The variance $\sigma^2_{\pmb{w}_\omega}$ captures the power of the noise in the random noise $\pmb{w}_\omega$ as

\begin{equation} \label{SigmaOmeg}
\mathbb{E}[\pmb{w}_\omega(t)\pmb{w}_\omega(t')^T]=\sigma^2_{\pmb{w}_\omega}\pmb{I}_{3\times3}\delta(t-t')
\end{equation}

The bias $\pmb{b}_{\omega}$ is modeled as a first-order Markov processes given by

\begin{equation} \label{OmegaBias2}
\dot{\pmb{b}}_{\omega}=\frac{-\pmb{b}_{\omega}}{\tau_\omega}+\pmb{w}_{b\omega}
\end{equation}

The variance $\sigma^2_{b\omega}$ captures the power of the noise in the random noise $\pmb{w}_{b\omega}$.

\subsection{Star tracker model}

In $\tilde{\pmb{q}}$, Gaussian white noise $\pmb{w}_{q}$, bias $\pmb{b}_q$ are included as the following

\begin{equation} \label{StarTrackerMeasure}
\tilde{\pmb{q}}=\partial \pmb{q}(\pmb{w}_{q})\otimes\partial \pmb{q}(\pmb{b}_q)\otimes\pmb{q}
\end{equation}

The variance $\sigma^2_{\pmb{w}_q}$ captures the power of the noise in the random noise $\pmb{w}_q$.

The bias $\pmb{b}_q$ is modeled as a first-order Markov processes given by

\begin{equation} \label{StartrackerBias}
\dot{\pmb{b}}_{q}=\frac{-\pmb{b}_{q}}{\tau_q}+\pmb{w}_{bq}
\end{equation}

The variance $\sigma^2_{\pmb{w}_{bq}}$ captures the power of the noise in the random noise $\pmb{w}_{bq}$.

The small rotation caused by $\partial \pmb{q}(\pmb{w}_{q})$ and $\partial \pmb{q}(\pmb{b}_q)$ is obtained by 

\begin{equation} \label{SmallRotationQQ}
\partial \pmb{q}\approx\begin{bmatrix}
\pmb{\theta}/2 \\ 1
\end{bmatrix}
\end{equation}

\subsection{Accelerometer model}

In $\tilde{\ddot{\pmb{r}}}$, Gaussian white noise $\pmb{w}_r$ and bias $\pmb{b}_r$ are included as the following

\begin{equation} \label{AccelerometerMeasure}
\tilde{\ddot{\pmb{r}}}=\ddot{\pmb{r}}+\pmb{b}_r+\pmb{w}_r
\end{equation}

The variance $\sigma^2_{\pmb{w}_r}$ captures the power of the noise in the random noise $\pmb{w}_r$.

The bias $\pmb{b}_r$ is modeled as a first-order Markov processes given by

\begin{equation} \label{OmegaBiasr}
\dot{\pmb{b}}_{r}=\frac{-\pmb{b}_{r}}{\tau_r}+\pmb{w}_{br}
\end{equation}

The variance $\sigma^2_{\pmb{w}_br}$ captures the power of the noise in the random noise $\pmb{w}_br$.

\subsection{Radio ranging sensor}
The radio ranging sensor is mounted on the follower spacecraft.

The desired relative distance between leader and follower is given by 
\begin{equation} \label{rDesired}
\pmb{r}^d_{rel}=\begin{bmatrix}
   0 & 0 & 1 \text{km}
\end{bmatrix}^T
\end{equation}

The measured $r^z_{rel}$ in the z axis is denoted as $\tilde{r}^z_{rel}$, and it includes misalignment $\epsilon_{rel}$ and white Gaussian noise given by

\begin{equation} \label{RadioMeasure}
\tilde{r}^z_{rel}=r^z_{rel}+\epsilon_{r^z}+w_{r^z}
\end{equation}

The misalignment $\epsilon_{r^z}$ is modeled as a first-order Markov processes given by

\begin{equation} \label{RadioRangeBias}
\dot{\epsilon}_{r^z}=\frac{-\epsilon_{r^z}}{\tau_{r^z}}+w_{\epsilon r^z}
\end{equation}

The variance $\sigma^2_{w_{\epsilon r^z}}$ captures the power of the noise in the random noise $w_{\epsilon r^z}$.

\subsection{Extended Kalman filter}
The residual error (also called the true navigation errors) is defined as

\begin{equation} \label{resi}
\pmb{\varepsilon}=\pmb{x}(t)-\hat{\pmb{x}}(t)
\end{equation}

Extended Kalman filter (EKF) \cite{markley2014fundamentals} is used for state estimation and sensor fusion for nonlinear dynamics. In EKF, the covariance of the residual error $\pmb{P}$ is used for updating the states. The first-order Markov process bias and misalignment in the sensors are considered as the states to be estimates by EKF. $\pmb{w}$ and $\pmb{\nu}_k$ Gaussian white noise with the covariance $\pmb{Q}$ and $\pmb{R}_k$ are considered for the dynamics model and sensor model, respectively. The subscript $k$ shows at time instant $t_k$. $x^+$ shows the updated value of $x$ after the measurement by EKF from the measurements, and $x^-$ shows the value of $x$ before the measurement. $x^-$ comes from the propagation.

EKF is given by the following equations

\textbf{Nonlinear dynamics model}
\begin{equation} \label{modelx}
\dot{\pmb{x}}=\pmb{f}(\pmb{x}(t),\pmb{\tau}(t),\pmb{w}, t)
\end{equation}

\textbf{Sensor model}
\begin{equation} \label{modely}
\pmb{y}_k=\pmb{h}(\pmb{x}_k)+ \pmb{\nu}_k
\end{equation}

\textbf{Initialization of states}
\begin{equation} \label{SysINI}
\hat{\pmb{x}}(t_0)=\hat{\pmb{x}}_0
\end{equation}

\textbf{Initialization of residual error covariance}
\begin{equation} \label{Covp}
\hat{\pmb{P}}(t_0)=\hat{\pmb{P}}_0
\end{equation}

\textbf{States propagation}
\begin{equation} \label{CovS1}
\hat{\dot{\pmb{x}}}=\pmb{f}(\pmb{x}(t),\pmb{\tau}(t), t)
\end{equation}

\textbf{The covariance of residual error propagation}
\begin{equation} \label{CovpPE}
\hat{\dot{\pmb{P}}}=\pmb{F}\pmb{P}+\pmb{P}\pmb{F}^T+\pmb{G}\pmb{Q}\pmb{G}^T
\end{equation}
\begin{equation*} \label{Covp1}
\pmb{F}=\frac{\partial{\pmb{f}}}{\partial{\pmb{x}}}|_{\hat{\pmb{x}}}
\end{equation*}
\begin{equation*} \label{Covp2}
\pmb{G}=\frac{\partial{\pmb{f}}}{\partial{\pmb{w}}}|_{\hat{\pmb{x}}}
\end{equation*}

\textbf{Updating EKF gain}
\begin{equation} \label{kupdate}
\pmb{K}_k=\pmb{P}^-_k\pmb{H}^T_k[\pmb{H}_k\pmb{P}^-_k\pmb{H}^T_k+\pmb{R}_k]^-1
\end{equation}
\begin{equation*} \label{kupdate2}
\pmb{H}_k=\frac{\partial{\pmb{h}}}{\partial{\pmb{x}}}|_{\hat{\pmb{x}}^-_k}
\end{equation*}

\textbf{Updating the states}
\begin{equation} \label{Stateupdate}
\hat{\pmb{x}}^+_k=\hat{\pmb{x}}^-_k+\pmb{K}_k(\pmb{y}_k-\pmb{h}(\hat{\pmb{x}}^-_k))
\end{equation}

\textbf{Updating the covariance of residual error propagation}
\begin{equation} \label{Pupdate}
\hat{\pmb{P}}^+_k=[1-\pmb{K}_k\pmb{H}_k]\hat{\pmb{P}}^-_k
\end{equation}

\section{Keplerian elements}\label{Keplerian}

The orbits are highly-elliptical supersynchronous geostationary transfer orbit with a 32.5-hour period. 
For the simulation, the Keplerian elements are eccentricity $\gamma$, semi-major axis $a$, inclination $i$, right ascension of the ascending node $\Omega$, and argument of periapsis $\omega$. The Keplerian elements are given in Table \ref{Keplerian elements}. 

\begin{table}[h!] \centering
\caption{Keplerian elements for leader and follower.}
\label{Keplerian elements} 
\setlength{\tabcolsep}{3pt}
\begin{tabular}{|p{80pt}|p{45pt}|p{45pt}|p{45pt}|p{55pt}|p{55pt}|}
\hline
 & $i$ deg & $\Omega$ deg & $\omega$ deg & $a$ km & $\gamma$
\\
\hline
leader orbit & 0.34 & 0 & 4.6743 & 45300 & 0.7125 \\\hline

follower orbit & 0.34 & 0 & 4.6743 & 45300 & 0.7336 \\\hline

\end{tabular}
\end{table}

\section{Spacecraft and instrumentation parameters}

The follower and leader are 6U CubeSat with inertial mass 10.2kg and nominal inertial momentum matrix $\bar{\pmb{J}}$ as
\begin{equation} \label{Inertia}
\bar{\pmb{J}}=\begin{bmatrix}
  0.1383&0&0 \\0&0.1577&0\\0&0&0.1039
\end{bmatrix}
\text{kg $\cdot$ m$^2$}
\end{equation}

The time variant modeling error $\delta \pmb{J}$ in $\pmb{J}$ \eqref{Mom} for both leader and follower is modeled as

\begin{equation} \label{DJ}
\delta \pmb{J}=
\begin{bmatrix}
  0.0038\sin(t) & 0 & 0 \\ 0 & 0.005\cos(t) & 0\\0 & 0 & 0.0011 
\end{bmatrix}\text{kg $\cdot$ m$^2$}
\end{equation}

The reaction wheel and thruster parameters for both leader and follower are given in Table \ref{RWParam}.

\begin{table}[h!] \centering
\caption{Reaction wheel and thruster parameters per axis.}
\label{RWParam} 
\centering
\setlength{\tabcolsep}{3pt}
\begin{tabular}{|p{100pt}|p{135pt}|p{135pt}|}
\hline
 & Leader 3-sigma value & Follower 3-sigma value
\\
\hline

$f_\tau$ per axis & 0.01 & 0.01 \\\hline

$\epsilon_\tau$ (mrad)/axis  & 1 & 1 \\\hline

$\pmb{w}_\tau \ (\text{N}-\text{m})$/axis & 0.001 & 0.001   \\\hline

$\pmb{w}_{b \tau} \ (\text{N}-\text{m})$/axis & 0.001 & 0.001  \\\hline

$\tau_\tau$ hr/cycle/axis & 1 & 1  \\\hline

$b_u $ (mN-m)/axis & 0.1 & 0.1 \\\hline

$f_u$ per axis & 0.01 & 0.01 \\\hline

$\epsilon_u$ (mrad)/axis  & 1 & 1 \\\hline

$\pmb{w}_u \ (\text{N})$/axis & 0.001 & 0.001   \\\hline

$\pmb{w}_{b u} \ (\text{N})$/axis & 0.001 & 0.001  \\\hline

$\tau_u$ (hr/cycle)/axis & 1 & 1  \\\hline

\end{tabular}
\end{table}

The variance of external disturbances are given in Table \ref{DisExter}. 

\begin{table}[h!] \centering
\caption{External disturbances.}
\label{DisExter} 
\centering
\setlength{\tabcolsep}{3pt}
\begin{tabular}{|p{100pt}|p{105pt}|p{115pt}|}
\hline
 & leader 3-sigma value & follower 3-sigma value
\\
\hline
$\pmb{w}_{\dot{\pmb{\omega}}} \ (\text{N}-\text{m})$/axis & 0.001 & 0.001   \\\hline

$\tau_g$ (hr/cycle)/axis & 1 & 1 \\\hline

$\pmb{w}_g$ micor-N/axis & 22.5 & 22.5 \\\hline

\end{tabular}
\end{table}

The sensor parameters for the IMU and star tracker are given in Table \ref{Sensors}. The radio ranging is mounted on the follower giving the relative distance between the follower and leader. The laser beacons are mounted the leader forming an interferometry sensor. The precise location of the leader is obtained with respect to the follower using the interferometry sensor. The sensor parameters for the interferometry sensor and radio ranging are given in Table \ref{SensorsIR}.

\begin{table}[h!] \centering
\caption{Sensor parameters.}
\label{Sensors} 
\centering
\setlength{\tabcolsep}{3pt}
\begin{tabular}{|p{80pt}|p{105pt}|p{119pt}|p{112pt}|}
\hline
Sensor & Symbol & Follower 3-sigma value & Leader 3-sigma value
\\
\hline

Gyro & $f_\omega$ per axis & 0.0003 & 0.0003 \\\hline

Gyro &  $\epsilon_\omega$ (mrad)/axis  & 3 & 3 \\\hline

Gyro & $\pmb{w}_\omega$ micor-N/axis & 22.5 & 22.5 \\\hline

Gyro & $b_\omega$ deg/hr/axis & 3 & 3 \\\hline


Star tracker & $\pmb{w}_{q}$ per axis & 41 milliarcsecond & 3 arcminute \\\hline

Star tracker & $\pmb{w}_{bq}$ per axis & 0.1 milliarcsecond & 3 arcminute \\\hline

Star tracker & $\tau_q$ (hr/cycle)/axis & 1 & 1 \\\hline

Accelerometer & $\pmb{w}_r$ nano-g & 1 & 1 \\\hline

Accelerometer & $\pmb{w}_{br}$ micor-N/axis & 1 & 1 \\\hline

Accelerometer & $\tau_r$ (hr/cycle)/axis & 1 & 1 \\\hline

\end{tabular}
\end{table}

\begin{table}[h!] \centering
\caption{Interferometry and radio ranging sensors' parameters.}
\label{SensorsIR} 
\centering
\setlength{\tabcolsep}{3pt}
\begin{tabular}{|p{120pt}|p{115pt}|p{105pt}|}
\hline
Sensor & Symbol & 3-sigma value 
\\
\hline

Radio ranging & $w_{r^z}$ m/axis & 1 \\\hline

Radio ranging & $w_{\epsilon r^z}$ m/axis & 1 \\\hline

Radio ranging & $\tau_{r^z}$ (hr/cycle)/axis & 1 \\\hline

Interferometry sensor  & $\pmb{w}_{{\pmb{r}}^{xy}}$ millimeter/axis & 0.2 \\\hline

Interferometry sensor & $\pmb{w}_{\pmb{\epsilon}{\pmb{r}}^{xy}}$ millimeter/axis & 0.01 \\\hline

Interferometry sensor & $\tau_{{\pmb{r}}^{xy}}$ (hr/cycle)/axis & 1 \\\hline

\end{tabular}
\end{table}

\bibliographystyle{elsarticle-num}  
\bibliography{ConOps}               

\end{document}